\documentclass{achemso}

\usepackage[version=3]{mhchem} 
\usepackage[dvipsnames]{xcolor}
 
\usepackage{tikz}
\usetikzlibrary{positioning}
 
\usepackage{stackengine}

\title
  {Role of Oxygen during Methane Oxidation on Pd1/PdO1\text{@}CeO$_2$ Surface: A Combined Density Functional Theory, Microkinetic, and Machine Learning Approach}
\author{Shalini Tomar}
\affiliation{Indo-Korea Science and Technology Center (IKST), Bangalore, India}

\author{Hojin Jeong}
\affiliation{Korea Institute of Materials Science (KIMS), South Korea}

\author{Joon Hwan Choi}
\affiliation{Korea Institute of Materials Science (KIMS), South Korea}

\author{Seung-Cheol Lee}
\affiliation{Indo-Korea Science and Technology Center (IKST), Bangalore, India}
\altaffiliation{Electronic Materials Research Center, Korea Institute of Science $\&$ Technology, Korea}

\author{Satadeep Bhattacharjee}
\affiliation{Indo-Korea Science and Technology Center (IKST), Bangalore, India}

\email{jchoi@kims.re.kr,  leesc@kist.re.kr,  s.bhattacharjee@ikst.res.in}
\begin{document}






\begin{abstract}
This work explores the role of oxygen in the industrial-based methane oxidation process for its catalytic decomposition. Oxygen, as a well-known oxidizing agent, plays a pivotal role in methane oxidation by facilitating the conversion of methane (CH$_4$) into carbon dioxide (CO$_2$) and water (H$_2$O). We report, how oxygen influences methane oxidation on single Pd and PdO clusters supported on the CeO$_2$(111) surface. Oxygen is introduced through two distinct mechanisms: (1) directly as part of palladium oxide (PdO), and (2) indirectly through the interaction of oxygen with a single Pd atom, which forms PdO$_x$ clusters on the CeO$_2$(111) surface. 
Through DFT calculations, we explored several reaction pathways of methane oxidation on the  
Pd1/PdO1\text{@}CeO$_2$(111) surface, and both Pd \& PdO single clusters were found to thermodynamically favor this process. 
The DFT calculated activation barrier for methane activation is 0.63 eV on PdO1\text{@}CeO$_2$(111). Notably, our results also highlight the crucial role of a single Pd atom in oxygen dissociation, which facilitates the formation of PdO$_2$ species. The presence of oxygen significantly lowers the activation barrier for methane oxidation by 0.36 eV, thereby improving the catalytic efficiency compared to oxygen-deficient conditions. We further explored the reaction selectivity, coverage-dependent production rates, degree of rate control, and turnover frequency using microkinetic modeling. The analysis reveals that the conversion of CH$_4$ into CO$_2$ and H$_2$O predominantly occurs at high temperatures. The rate constants for various reaction steps were derived using the Sure Independence Screening and Sparsifying Operator (SISSO) method, a machine learning-based symbolic regression approach. This allowed us to build a predictive model for the rate constants, leveraging descriptors such as charge, coordination number, and interatomic distances, providing deeper insight into the factors influencing catalytic activity. These findings offer valuable guidance for optimizing catalysts for methane oxidation, a reaction of significant environmental and industrial importance.
\end{abstract}

\section{Introduction}
Methane (CH$_4$) is a significantly more potent greenhouse gas than carbon dioxide (CO$_2$)\cite{Zhan, Maxim} and to reduce the methane emission, catalytic decomposition of unburned CH$_4$ is essential \cite{Rouet, Mao, jean}. The conversion of methane is challenging due to its substantial energy barrier, which arises from the molecule's high symmetry and the strong interactions of its C–H bonds \cite{Hongyan, Burch, norskov}. 
A widely recognized approach for methane decomposition is by using oxygen, a well-known oxidizing agent, which facilitates the conversion of methane into CO$_2$ and H$_2$O, commonly referred to as methane oxidation \cite{havran, webley}. For the catalytic decomposition of methane, palladium-based catalysts stand out as the most effective metal catalyst among noble metals\cite{Yin, xiansheng, teng, zhu, willis}. The reported methane activation barrier for PdO(100) and Pd(111) surfaces ranges from 1.0 eV to 1.6 eV \cite{kinnunen2011,senftle2015,HERRON2012, Mikkel2016}. The nanoclusters of noble metal often exhibit exceptional catalytic activity due to the coordinatively unsaturated nature of individual clusters, which enhances the activation of reactants \cite{kumar, Sun, Feng, Wu}. 
Additionally, methane activation on pure Transition-metal (TM) surfaces, including noble metals such as Pt, Ru, Rh, and Pd, is notably challenging due to their low reactivity. TM-based catalysts supported on oxides such as cerium-based oxides (CeO$_2$) gained attention for their cost-effectiveness and enhanced performance for methane oxidation reactions\cite{Akri2019, montini2016}. 
A DFT study on a palladium cluster supported on CeO$_2$ demonstrates that the Pd cluster changes its oxidation states, thereby reducing the activation barrier \cite{senftle2017methane}. In addition to it, several theoretical studies reported the methane activation barrier on the TM supported CeO$_2$ less than 1 eV \cite{pablo2016,pablo2021, wang2018}. Mao et. al. conducted a comprehensive study on methane oxidation using palladium nanoparticles featuring pristine and oxygen-coated surfaces and they reported that during the catalytic process of methane oxidation, palladium particles undergo partial oxidation to form PdO \cite{QianMao, RobertF}. A recent experimental study revealed that during the methane oxidation, the likelihood of Pd metal transitioning to PdO increases with a higher CH$_4$/O$_2$ ratio or a decrease in temperature, leading to a substantial enhancement in the catalytic activity for methane oxidation\cite{shi, Abhaya}. Iglesia et. al. have demonstrated that turnover rates increase as Pd clusters undergo bulk oxidation, forming PdO clusters \cite{Iglesia}. 

Moreover, the catalytic activity of these Pd-based catalysts is highly dependent on the nature of the support material, the oxidation state, and the stability of palladium species and their interaction with the support and environmental elements \cite{Chen, wang1, Du, Petrov}. Oxygen is one of the co-reactants of CH$_4$ and plays a crucial role in its complete oxidation as an environmental element\cite{zhangten}. Importantly, the presence of both Pd and PdO on the particle surface serves to lower the activation energy required for methane oxidation effectively but there is still debate about which is better of them \cite{Bunting, Xiong, Yue}. Some studies suggest that metallic Pd is more effective than PdO in the complete oxidation of methane \cite{Hellman, Yue}. However, recent research points to PdO$_x$ or a metal/oxide interface as the key to catalytic activity\cite{stotz}. These differing conclusions may arise from the dynamic coexistence of Pd and PdO during reactions, making it difficult to pinpoint specific active structures and establish clear structure-activity relationships. PdO nanostructure shows good stability with the metal oxide surfaces such as CO$_3$O$_4$ \cite{carmen, Koo}, CeO$_2$ \cite{Dai}, NiO \cite{Wang}, MgO\cite{chen1}, ZnO\cite{young, Mhlongo}, Al$_2$O$_3$ \cite{Jon, xiang}, graphene oxide \cite{zheng} and zeolite \cite{wang2}. Wolf et. al. reported that the PdO/CeO$_2$ nanostructure exhibits high catalytic activity for low-temperature methane oxidation and the presence of water significantly reduces the activity \cite{Dai}. Some experiments show that in excess presence of O$_2$, Pd single atoms are readily activated into PdO$_x$ single nanocluster which helps to boost the catalytic activity for methane oxidation \cite{Jiang2020}. 

Apart from this, understanding complex reaction mechanisms is essential for designing better catalysis in heterogeneous catalysis. While recent advances in in-situ experiment techniques have improved our ability to study the catalytic reaction, tracking complete reaction pathways remains challenging, especially when multiple competing mechanisms are involved. In such cases, first principle calculations paired with microkinetic modeling offer a valuable approach to exploring reaction kinetics \cite{bossche, Kulkarni, Motagamwala, keller}. In this study, we explored the complete reaction pathways of a single PdO$_x$ cluster supported on the Ceo$_2$(111) surface for methane oxidation using density functional theory (DFT). The CeO$_2$(111) surface was selected for its well-established stability and enhanced reactivity \cite{SGupta, sara202}. Additionally, we investigated the reaction path selectivity, production rate, degree of rate control, and turnover frequency through microkinetic modeling.

\section{Computational Methodology}
\subsection{First-principles study}
This investigation employs first-principles calculations using density functional theory (DFT) \cite{DFT1, DFT2, kohnsham} within the Vienna Ab initio Simulation Package (VASP) \cite{VASP1}. The Perdew-Burke-Ernzerhof (PBE) method is chosen for exchange-correlation energy, while the projected augmented wave (PAW) method \cite{PAW} describes the core-valence electron interaction. A plane wave basis set with an energy cutoff of 500 eV is employed, and the Brillouin zone is sampled using a $4 \times 4 \times 1$ k-points mesh. A U value of 5 eV is applied to the Ce 
f-orbitals, as used in earlier theoretical studies \cite{Dudarev}. A  dditionally, we incorporate the semi-empirical D3 van der Waals (vdW) correction to enhance the description of the adsorbed system \cite{DFTD}, and charge transfer is determined using the Bader charge analysis method \cite{bader}. The study focuses on the (111) surface of CeO$_2$, known for its good reactivity and stability\cite{penschke, lawler, Fan}. The CeO$_2$(111) surface is modeled with a $1 \times 1 \times 1$ conventional unit cell consisting of 4 layers (12 O-Ce-O atomic sublayers with 16 CeO$_2$ units). Following the benchmark tests, we decided to use a $1 \times 1 \times 1$ conventional unit cell for this study, rather than a larger supercell, as outlined in our previous work \cite{tomar}. The lattice constant of this surface is 7.68 \AA~, with a thickness of 10.7 \AA~, which we used throughout our calculations, and an additional 18 \AA~ vacuum along the z-direction to minimize interactions between periodic images. The bottom six atomic layers are fixed, while the upper six layers can relax until the maximum forces on each atom are below 0.005 eV/Å. To identify the most stable sites for the adsorption of single Pd and PdO, the adsorption energy (E$_{ads}$) is calculated using the formula; E$_{ads}$ = E$_{\text{surface + adsorbate}}$ - E$_{\text{surface}}$ - E$_{\text{adsorbates}}$, where each term refers to the DFT calculated total energy of the respective system. The thermodynamic behavior of each intermediate reaction step during methane oxidation is analyzed by computing the enthalpy change ($\Delta H$) as the difference between the DFT-calculated total energies of the product and reactant: $\Delta H$ = E$_{\text{product}}$ - E$_{\text{reactant}}$. The activation barrier (E$_a$) is determined using the climbing image nudged elastic band (CI-NEB) method \cite{NEB}, which is considered as the difference between the total energy of the transition state (TS) and the initial state (IS); E$_a$ = E$_{\text{TS}}$ - E$_{\text{IS}}$. Additionally, we developed a Brønsted–Evans–Polanyi (BEP) relationship to estimate the reaction barriers of intermediate steps involved in methane oxidation (see Supporting Information).

\subsection{Microkinetic Modelling}
The microkinetic modeling was performed using the MKMCXX software \cite{Filot}, with the assumption that all adsorbed species occupy a single position. For surface reactions,  the rate constant of the forward and reverse reactions can be derived from the Eyring equations, k = $\frac{K_BT}{h}   \exp{(\frac{- \Delta G^\ddagger}{R T}})$, where, k is the reaction rate constant, $\Delta$ G$^\ddagger$ is the Gibbs free energy, R is the gas constant and T is the temperature in Kelvin, respectively. The following equation is used to calculate the adsorption and desorption rates,

\begin{equation}
   K_{ads}  = \frac{PA}{\sqrt{2\pi mK_bT}}S 
\end{equation}

and the 

\begin{equation}
   K_{des}  = \frac{K_bT^3}{h^3}   \frac{A(2\pi mK_b)}{\sigma \theta}e^{-E_{des}/K_bT}
\end{equation}

Where, P, A, m, S, $\sigma$, $\theta$ and E$_{des}$ denote the partial pressure, the surface area of the adsorption site, relative molecular mass, the sticking coefficient, the symmetric number, rotational characteristics temperature and the desorption energy, respectively. The value of symmetry number ($\sigma$) for CH$_4$, O$_2$, CO$_2$, H$_2$O is 12, 2, 2, 2 \cite{donald} and rotational temperature ($\theta$) for CH$_4$, O$_2$, CO$_2$, H$_2$O is 7.54 K, 1.47 K, 0.56 K, 20.9 K, respectively \cite{atkins, adriana, donald}. To simplify our simulations, the sticking coefficients were set to 1 throughout all microkinetic simulations \cite{Jiao, Jianlin}. The apparent activation energy can be calculated from the equation 
\begin{equation}
  r =  k_{ads}.p_A -k_{des}.[A_{ads}]
\end{equation}

\begin{equation}
  E_{app} = RT^2 \frac{d \ln r}{dT} = -R\frac{d \ln r}{d \frac{1}{T}}  
\end{equation}

\begin{equation}
  X_{c,i} = \frac{k_i}{r_c} (\frac{\partial r_c}{\partial k_i})_{}k_j = (\frac{\partial \ln r_c}{ \partial \ln k_i})
\end{equation}

where r is the total rate of the equation and T is the temperature. The degree of rate control (DRC) refers to the relative change of the rate caused by the rate constant of a given elementary reaction step while keeping the equilibrium constant.


\section{Results \& Discussion}
This work is divided into two main sections. First, we analyzed the complete reaction pathways using DFT, followed by a discussion on production rates, turnover frequency, and the degree of rate control based on microkinetic modeling. The DFT analysis is further divided into two parts: investigating the direct and indirect effects of oxygen through Pd and PdO supported by CeO$_2$(111). In the first part, we examined the interaction between PdO and CeO$_2$ (direct effect). In the second part, we studied how O$_2$ interacts with a single Pd atom to form PdO$_x$ species (indirect effect). The complete reaction pathways for Pd1\text{@}CeO$_2$ under dry conditions have already been investigated by our group\cite{tomar}. In this study, we focus specifically on the reaction pathway for Pd1\text{@}CeO$_2$ under oxygen-rich conditions.

\subsection{Adsorption of Pd and PdO over CeO$_2$(111)}
\begin{figure*}
\centering
{\includegraphics[trim=1.0cm 2.0cm 0.80cm 1.2cm,clip=true,width=0.8\textwidth]{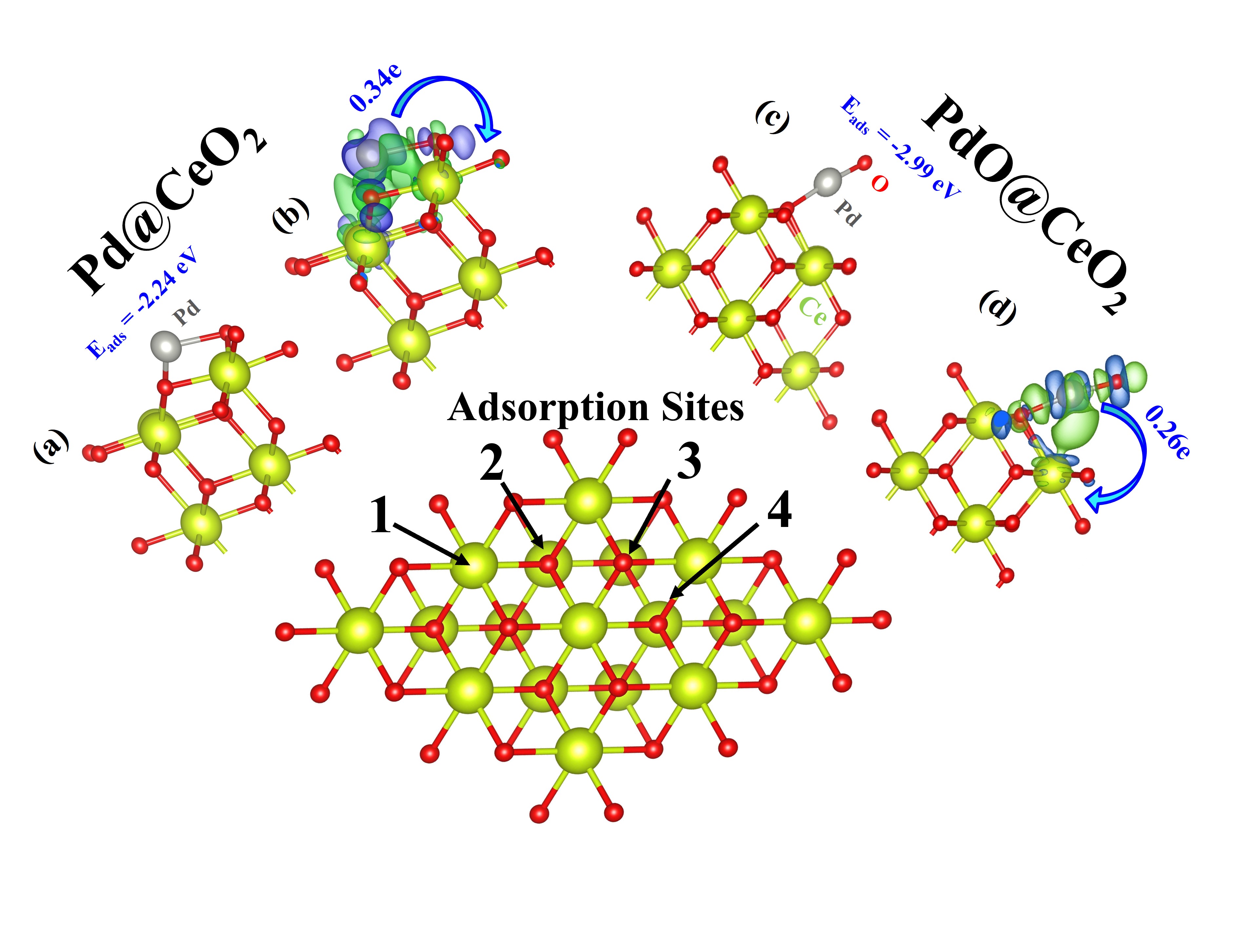}}
\caption{The adsorption site for single Pd and PdO over the optimized CeO$_2$ supercell (in the middle), where 1, 2, 3 and 4 refers to the top of Ce atom, top of oxygen, top of hexagon and bridge of Ce \& O. The relaxed geometry/charge transfer of Pd1\text{@}CeO$_2$ and PdO1\text{@}CeO$_2$ in (a)/(b) and (c)/(d), respectively. The green and blue colors indicate the accumulation and depletion of electrons.}
\label{Fig1}
\end{figure*}

We chose four distinct sites, as depicted in {Figure~\ref{Fig1}} (in the middle), to study the adsorption of Pd and PdO on the CeO$_2$(111) surface. In addition to examining the adsorption sites, for PdO, we also investigated the orientation during adsorption by considering both Pd and O facing toward the surface. All possible adsorption sites and orientations of PdO have been depicted in SFig1. The adsorption energies of all possible sites and orientation for Pd and PdO adsorption are mentioned in the table.1 in the supporting information. For both Pd and PdO, the most stable adsorption site is the bridge site (site 4), with adsorption energies of -2.24 eV for Pd and -2.99 eV for PdO, indicating a stronger interaction of PdO with the surface compared to a single Pd atom. These adsorption energies agree with previously reported values \cite{Moraes, spezzati, tomar}. In case of PdO, the orientation of Pd towards the surface is the most stable geometry as compared to O towards the surface. The Pd of PdO is connected with O$_S$ on the surface and constructs PdO$_2$ local species.

The optimized Pd1\text{@}CeO$_2$ and PdO1\text{@}CeO$_2$ structures at the most stable site have been shown in {Figure~\ref{Fig1}} (a) \& (c), respectively. In our previous work, we reported the adsorption of a single Pd atom on the CeO$_2$ surface\cite{tomar}. In this study, we focus primarily on the adsorption of PdO on the CeO$_2$(111) surface. The variance in adsorption energy of PdO compared to the top of the oxygen site (site 2) and the top of the hexagon site (site 3) is 0.37 eV and 0.36 eV, respectively. Upon adsorption of PdO on the CeO$_2$ surface, the Pd-O bond length reduces to 1.79 \AA~, slightly shorter than that in the PdO molecule (1.81 \AA~), with a distance of 1.97 \AA~ from the surface.
The O atom of PdO is slightly bent towards the Ce atom at the surface. Similar to a single Pd atom, PdO also bonds with two oxygen atoms; one from the PdO molecule and the other from the CeO$_2$ surface, forming a PdO$_2$ local species by changing its oxidation state from Pd$^0$ to Pd$^{+2}$, which is more catalytically active.
Through the Bader charge analysis, it was confirmed that during the adsorption, both Pd and PdO transfer 0.34e and 0.26e charges to the surface, reducing Ce$^{4+}$ to Ce$^{3+}$ as shown in {Figure~\ref{Fig1}(b) \& (d)}.


\subsection{Intermediate Species in Associative and Dissociative form on PdO\text{@}CeO$_2$}

\begin{figure*}
\centering
{\includegraphics[trim=0.150cm 0.50cm 0.300cm 0.20cm,clip=true,width=1.00\textwidth]{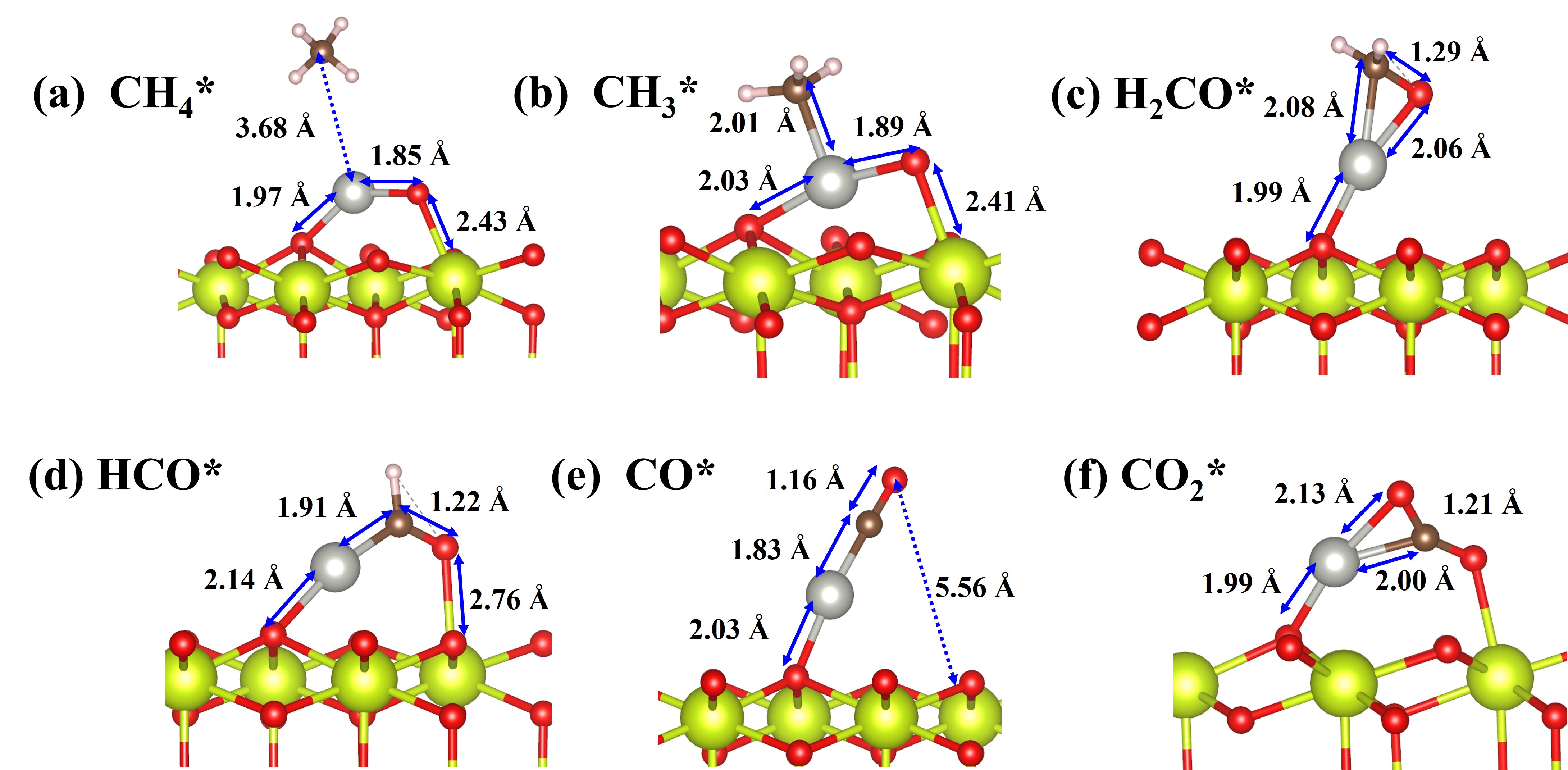}}
\caption{Geometrical representation of adsorbed carbon species in the associative form on PdO supported with CeO$_2$(111) surface; (a) CH$_4^*$, (b) CH$_3^*$, (c) H$_2CO^*$,  (d) HCO$^*$, (e)CO$^*$, and (f) CO$_2^*$.}
\label{Fig2}
\end{figure*}

When considering the adsorption of carbon species in its associative form on the PdO1\text{@}CeO$_2$(111) surface, there exists a competition between two sites: Pd and O as shown in {Figure~\ref{Fig1}(c)}. We compared the adsorption energy for each carbon species on both sites, and it was found that the 'Pd' site is thermodynamically more catalytically active. The adsorption energies for all carbon species on both the 'Pd' and 'O' active sites are provided in  {Table 1} in the supporting information.
The optimized structural geometries of the intermediate species adsorbed on PdO1\text{@}CeO$_2$ are illustrated in their associative form in {Figure~\ref{Fig2}} and in their dissociative form in SFig3

In the case of methane (CH$_4$) adsorption, methane is physisorbed over Pd1/CeO$_2$(111) in a scissor geometry, with an adsorption energy of -0.21 eV, similar for both sites. The adsorption of methane retains the surface geometry while pushing the PdO molecule closer to the surface, where the oxygen atom of PdO forms a bond with the Ce atom of the surface at a bond length of 2.43 \AA~ as shown in {Figure~\ref{Fig2}(a)}. The distance between methane and the Pd atom measures 3.68 \AA~. However, in the case of CH$_3$ adsorption, the adsorption energies are -2.03 eV and -2.97 eV for the Pd and O sites, respectively.
According to the Sabatier principle, the Pd site proves to be thermodynamically more favorable compared to the 'O' site. Additionally, the adsorption of CH$_3$ further suppresses PdO towards the surface, bonding with the Pd atom in an umbrella-style configuration, with a distance of 2.01 \AA~ between the CH$_3$ molecule and the Pd atom, which can be seen in {Figure~\ref{Fig2}(b)}. Following the adsorption of CH$_4$ and CH$_3$ species, there's a localized formation of PdO$_2$ species. Similar to CH$_3$, both CH$_2$ and CH species also exhibit a preference for the Pd site, displaying adsorption energies of -5.94 eV and -7.88 eV, respectively, which are notably lower than those of -0.20 eV and -1.30 eV observed for the O site. The adsorbed CH$_2$ species form bonds with both Pd and oxygen atoms, with the carbon positioned at a distance of 2.08 \AA~ from Pd and 1.29 \AA~ from oxygen. Similarly, CH also bonds with Pd and O atoms, with the carbon situated at distances of 1.91 \AA~ and 1.22 \AA~, respectively. The adsorption energy for single C and H is the same for both the Pd and O sites (See Table. 1, in supporting information).

\begin{figure*}
\centering
{\includegraphics[trim=0.6cm 0.0cm 0.6cm 0.0cm,clip=true,width=1.0\textwidth]{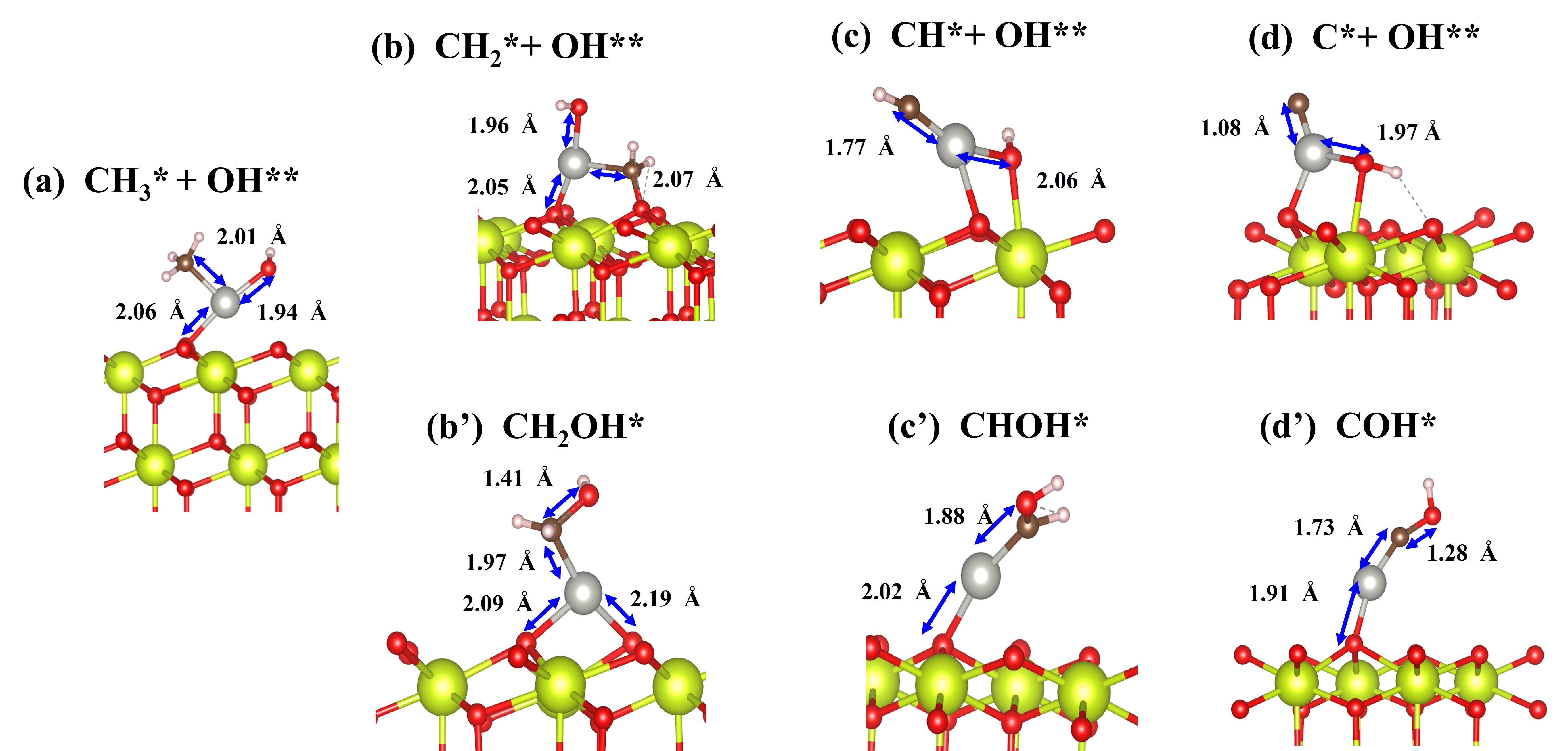}}
\caption{The optimized geometries of other intermediates over PdO\text{@}CeO$_2$ during methane oxidation. The '*' and '**' show the Pd and O as the adsorption sites. The (b), (c), (d) and (b'), (c'), (d') represent the intermediate species for path1 and path2.}
\label{Fig3}
\end{figure*}
In addition to the adsorption of species in their associative form, we also examined the intermediate species in their dissociative form during methane oxidation as shown in {Figure~\ref{Fig3}}, where 'Pd' and 'O' were considered as the active sites. Upon dissociation of CH$_4$, CH$_3^*$ bonds to Pd and H$^{**}$ bonds to O, where '*' and '**' represent their attachment to Pd and O, respectively. The bond distances are 2.01 \AA~ for CH$_3^*$ to Pd and 1.01 \AA~ for H$^{**}$ to O, which is comparable to the OH bond length. This confirms that CH$_4$ dissociates into CH$_3$ and OH (CH$_4$  $\rightarrow$  CH$_3^*$ + OH$^{**}$) as depicted in {Figure~\ref{Fig3}(a)}. During CH$_3$ dissociation, two stable intermediates were identified: CH$_2^*$ + OH$^{**}$ and CH$_2$OH$^*$. In both cases, Pd maintains the same coordination, bonding with two oxygen atoms and one carbon atom. These two stable intermediates proceed along two different reaction pathways with distinct intermediates: CH$_2^*$ + OH$^{**}$ $\rightarrow$ CH$^*$ + OH$^{**}$ $\rightarrow$ C$^*$ + OH$^{**}$ (in {Figure~\ref{Fig3}(b)-(d)}), and CH$_2$OH$^*$ $\rightarrow$ CHOH$^*$ $\rightarrow$ COH$^*$ (in {Figure~\ref{Fig3}(b')-(d')}). In the CH$^*$ + OH$^{**}$, the Pd atom forms a bond with CH$^*$ at a distance of 1.77 \AA~, while OH is 2.06 \AA~ away from Pd and in the CHOH$^*$, the distance between Pd and CHOH$^*$ is 1.88 \AA~, showing that Pd bonds more strongly with CH$^*$ compared to CHOH$^*$. Thus, during the dissociation of CH$^*$ to C$^*$ + OH$^{**}$ and CHOH$^*$ to COH$^*$, Pd exhibits coordination numbers of 3 and 2, respectively. The C$^*$ species bonds more strongly with Pd compared to COH$^*$.

\section{Reaction pathway for Methane Oxidation at PdO1\text{@}CeO$_2$}
Our primary focus was on the dissociation of methane into CO$_2$, and we investigated the reaction pathway of methane dissociation over the {PdO1\text{@}CeO$_2$(111) surface. As we found two active sites over the PdO1\text{@}CeO$_2$(111) surface, the active site O assumes a crucial role, exhibiting two distinct reaction pathways. In one pathway, it continues to act as an active site, while in the other, it undergoes a change in behavior, transitioning from an active site to an element of the product. Both reaction pathways are shown in {Figure~\ref{Fig4}} with the reaction energy of their respective reaction steps and the intermediates of these paths can be seen in {Figure~\ref{Fig5}}.  The thermodynamic behavior of both reaction pathways is different, and there are notable differences in their reaction energies. This discrepancy aids in determining the preferred reaction pathway for methane dissociation into CO$_2$.
\begin{figure*}
\centering
{\includegraphics[trim=0.0cm 0.0cm 0.0cm 0.00cm,clip=true,width=0.8\textwidth]{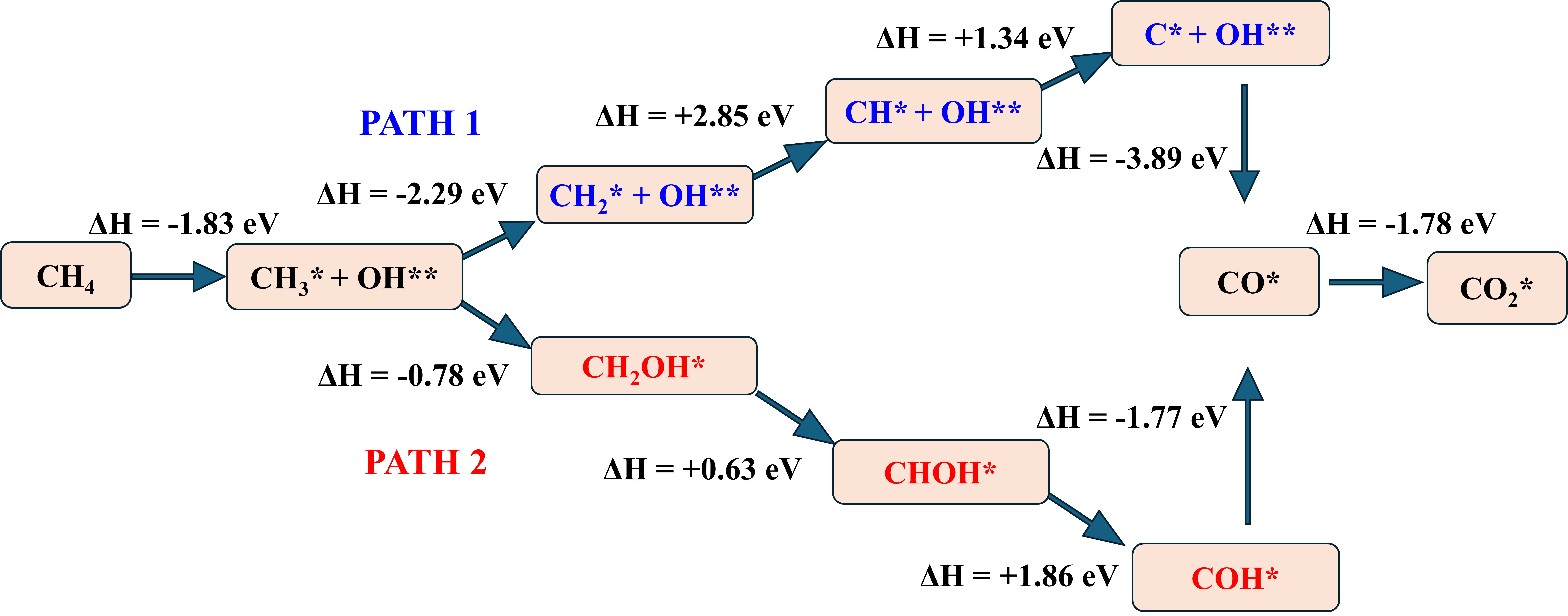}}
\caption{Two distinct reaction pathways with respective Gibbs free energy for methane conversion into CO$_2$ over PdO1\text{@}CeO$_2$(111). Some intermediates are common in both path 1 (blue color) and path 2 (red color).}
\label{Fig4}
\end{figure*}

The first step for both reaction pathways is similar as represented in {Figure~\ref{Fig4}}, where methane is initially physisorbed on the surface and undergoes an exothermic reaction with a reaction energy of -1.83 eV. Subsequently, the methane molecule breaks into CH$_3^*$ and H$^{**}$ species, with CH$_3^*$ bonding with Pd and H$^{**}$ species bonding with O. These bonding preferences correspond to the respective thermodynamically favorable sites discussed in the previous section. The required energy barrier to break CH$_4$ on PdO1\text{@}CeO$_2$ is 0.63 eV, calculated using a nudged elastic band method with DFT. In the second step of the reaction path, which involves removing the second hydrogen from the CH$_3^*$ species, two distinct products are formed CH$_2^*$+OH$^{**}$ and CH$_2$OH$^*$ in reaction path 1 and reaction path 2, respectively. During the formation of CH$_2^*$+OH$^{**}$, CH$_2^*$ is situated on the Pd active site while H$^{**}$ is located on the O active site, as depicted in the third box in path 1 in reaction pathways 1. This process exhibits thermodynamically exothermic behavior with a reaction energy of -2.29 eV. The third step CH$_2^*$  $\rightarrow$  CH$^*$ + OH$^{**}$ and fourth steps CH$^*$ $\rightarrow$  C$^*$ + OH$^{**}$ of reaction path 1 are thermodynamically endothermic with reaction energy +2.85 eV and +1.34 eV. 

\begin{figure*}
\centering
{\includegraphics[trim=0.0cm 0.0cm 0.0cm 0.00cm,clip=true,width=0.8\textwidth]{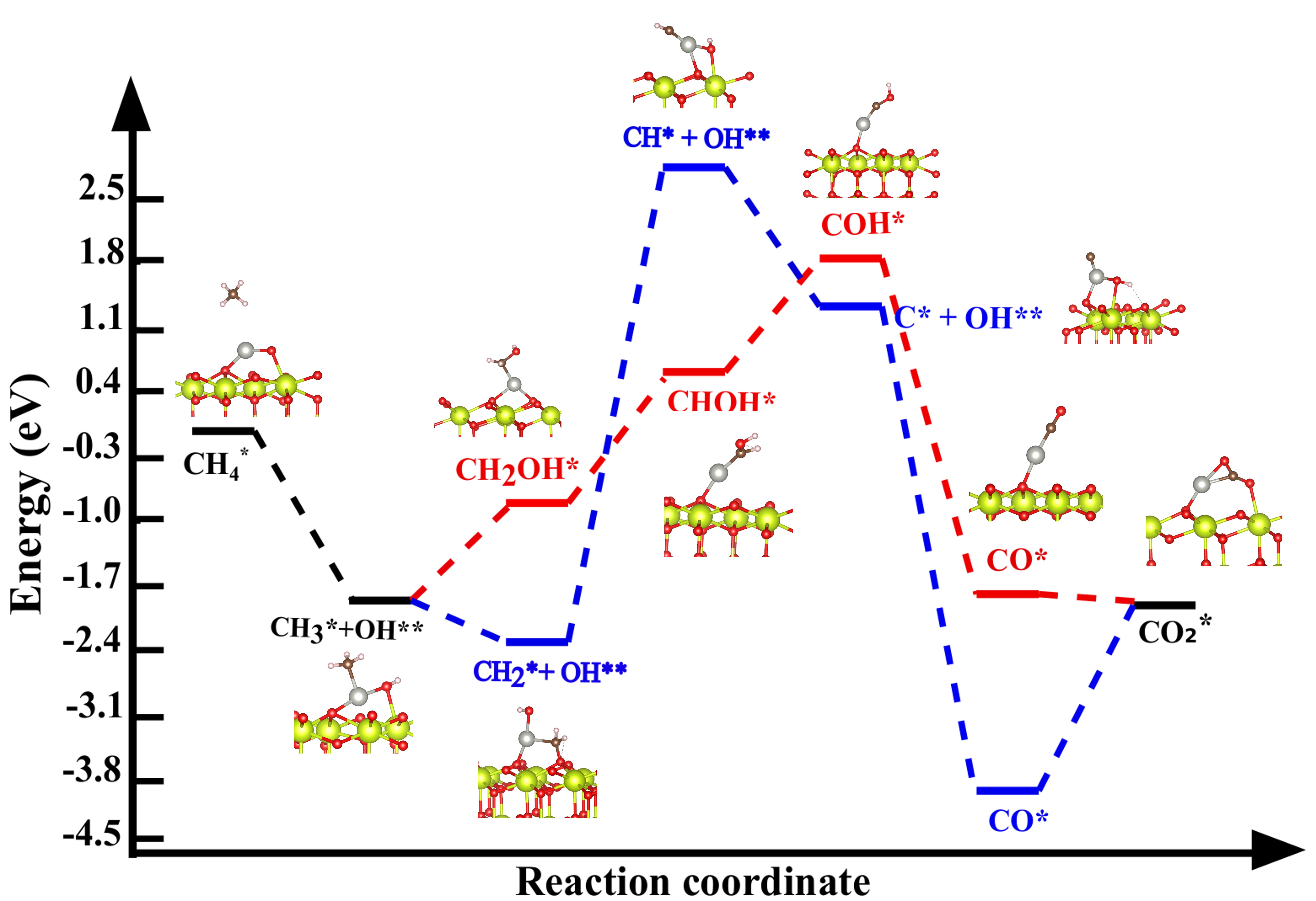}}
\caption{Complete reaction pathways for methane dissociation into CO$_2$ over PdO1\text{@}CeO$_2$(111). Blue and red colors refer to Path 1 and Path 2, whereas black color represents the common intermediates in Path 1 and Path 2. The '*' and '**' depicted the Pd and O as an active site.}
\label{Fig5}
\end{figure*}

In pathway 2 of the reaction, the energy barrier for the reaction is notably lower compared to pathway 1 due to the formation of a different product. In the second step of pathway 2, we observe the formation of alcohol after releasing 0.78 eV of energy, showing the exothermic behavior, whereas in the third step, the formation of alcohol occurs after absorbing +0.63 eV of energy, demonstrating a thermodynamically endothermic behavior.

\section{Reaction Pathway for Methane Oxidation at Pd1\text{@}CeO$_2$ in Oxygen-rich Condition}
In our work, we focused on the role of oxygen in both direct and indirect approaches during methane oxidation. The previous section focused on methane oxidation over a single PdO supported on the CeO$_2$(111) surface, specifically investigating the direct role of oxygen as PdO. In this section, we will discuss the indirect approach, where O$_2$ interacts with CH$_4$ to form a local PdO$_x$ geometry that serves as an active site. This method is referred to as methane oxidation in oxygen-rich conditions, as the dissociation of methane occurs in the presence of oxygen. In our earlier work, we already studied methane dissociation in the absence of oxygen, known as dry conditions or deep dehydrogenation \cite{tomar}. The complete reaction pathway for methane dissociation into CO$_2$ in the presence of oxygen is depicted in {Figure~\ref{Fig6}}, and a structural representation of this process can be seen in SFig.3. (See Figures S4, S5, and S6 in the supporting information for the possible configurations of intermediate species adsorbed on the Pd@CeO$_2$ surface).


\begin{figure*}
\centering
{\includegraphics[trim=0.150cm 0.50cm 0.300cm 0.20cm,clip=true,width=1.00\textwidth]{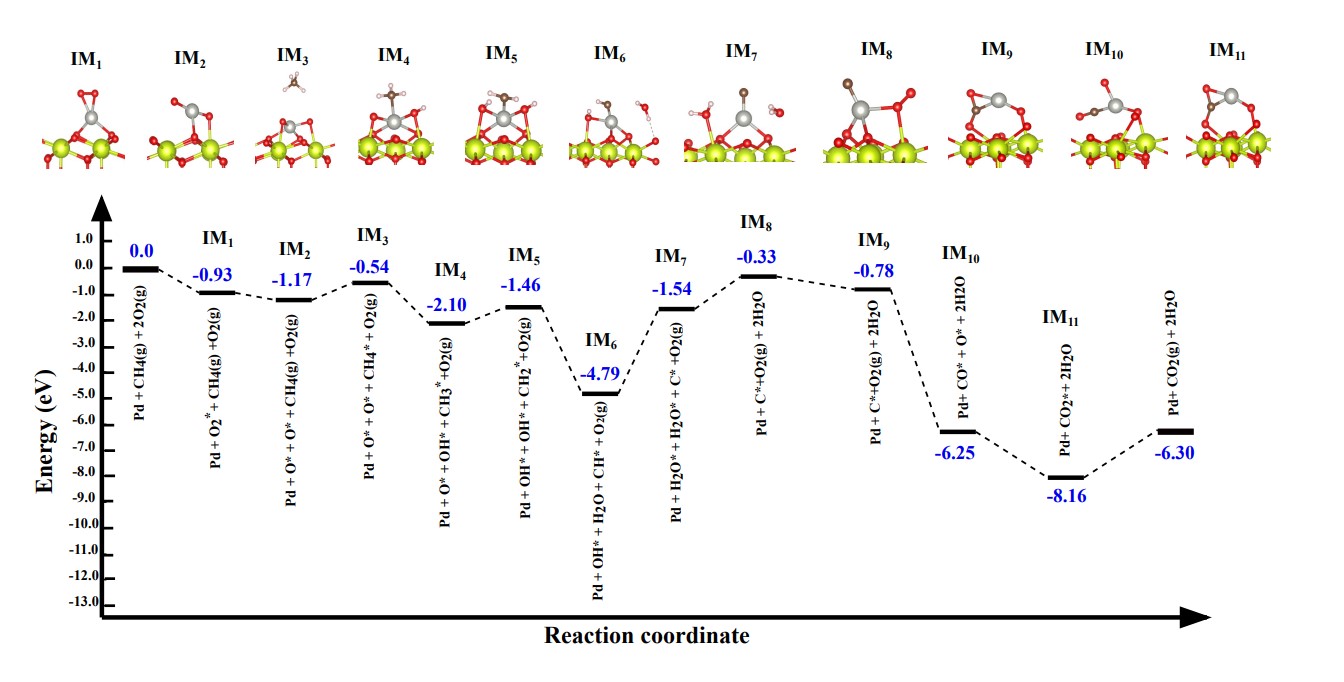}}
\caption{The complete reaction pathways of methane oxidation over Pd\text{@}CeO$_2$(111) in the presence of oxygen with the structural view of corresponding intermediates.}
\label{Fig6}
\end{figure*}

During our investigation of methane oxidation in oxygen-rich conditions, we found that CH$_4$ and O$_2$ compete for adsorption on the active site, as both molecules tend to bond with Pd. When O$_2$ adsorbs on Pd1\text{@}CeO$_2$, it binds symmetrically with Pd, forming localized PdO$_2$ species. The adsorption energy of the O$_2$ molecule is -1.63 eV, indicating a stronger binding with Pd compared to CH$_4$, which has an adsorption energy of -0.21 eV. The bond length between the oxygen atoms in O$_2$ extends from 1.20 \AA~ (in the isolated molecule) to 1.31 \AA~ (after adsorption over Pd1\text{@}CeO$_2$), while the distance of both oxygen atoms from Pd is 2.03 \AA~. Upon  O$_2$ adsorption on Pd1\text{@}CeO$_2$(111) surface, it undergoes an exothermic reaction, indicating thermodynamic favorability with a reaction energy of -0.92 eV. In the subsequent step, the adsorbed oxygen molecule dissociates into two individual O$^*$ atoms, which then act as active sites in the methane dissociation process. The dissociation of oxygen also is a thermodynamically exothermic reaction with an energy of -1.17 eV. The activation barrier for O$_2$ dissociation is 2.12 eV, which lies in the range of previous values\cite{zhang, shiming}. During the dissociation, the distance between the dissociated oxygen atoms increases from 1.31 \AA~ (before dissociation) to 2.87 \AA~, with one oxygen atom elongating towards the surface Ce atom. The bond length of the dissociated oxygen atoms from Pd is 1.79 \AA~ and 1.87 \AA~.
After the dissociation of O$_2$, the next step of the reaction path is the dissociation of methane, where dissociated O$^*$ acts as an active site for methane adsorption. The adsorption of CH$_4$ causes the PdO$_2$ species to be pushed toward the surface, while Pd forms additional bonds with two oxygen atoms from the surface. The methane molecule breaks down into CH$_3$ and H, where CH$_3^*$ binds to the Pd atom and H binds to the oxygen atom of O$_2$, forming the OH$^*$ species, as depicted in {Figure~\ref{Fig6}}. The dissociation of methane over Pd1\text{@}CeO$_2$(111) exhibits an exothermic nature thermodynamically with a reaction energy of -2.10 eV. 


Similarly, CH$_3^*$, CH$_2^*$, and CH$^*$ undergo stepwise dehydrogenation, leading to the sequential formation of OH$^*$ and H$_2$O$^*$ species. Once the water species is formed, it weakly bonds to the surface before desorbing, leaving behind the C$^*$ species, which can be seen in {Figure~\ref{Fig6}}. Once the water molecule is eliminated, the remaining carbon species obstruct the active Pd site. 
Subsequently, an O$_2$ molecule adsorbs onto the Pd surface and dissociates in the presence of the C$^*$ species. The dissociation of O$_2$ required for the subsequent formation of CO$_2$ as a product becomes significantly difficult. The presence of adsorbed C$^*$ on top of Pd resulted in a slowed dissociation process. Upon comparing the adsorption energies of O$_2$ molecules in the absence and presence of carbon species at the Pd site, we observed a significant weakening of the O$_2$ molecule's bonding in the presence of carbon species, leading to a reduction in adsorption energy by more than half. Specifically, in the absence of C$^*$ at Pd, the adsorption energy measured -0.89 eV and bond length 1.34 \AA~, which is elongated as compared to isolated O$_2$ molecule. While in the presence of C$^*$ at Pd, the adsorption energy of the O$_2$ molecule decreased to -0.42 eV and the bond length was 1.29 \AA~ which is almost similar to the isolated molecule which concludes that in the presence of C$^*$, the O$_2$ molecule is less activated for the dissociation. In summary, the steps involved in methane oxidation can be described as follows, (i) the adsorption of methane on Pd1\text{@}CeO$_2$ surface in the presence of O$_2$; {\footnotesize(Pd+CH$_4$(g) $\rightarrow$ Pd+CH$_4^*$)}; (ii) removing of hydrogen step by step from CH$_4^*$, CH$_3^*$, CH$_2^*$,  and CH$^*$ species {\footnotesize (CH$_4^*$ $\rightarrow$ O$^*$+OH$^*$+CH$_3^*$ $\rightarrow$ OH$^*$+OH$^*$+CH$_2^*$ $\rightarrow$  O$^*$+OH$^*$+CH$^*$ $\rightarrow$ H$_2$O$^*$+H$_2$O$^*$+C$^*$)}; (iii) desorption of H$_2$O and CO$_2$ molecules {\footnotesize (H$_2$O$^*$+H$_2$O$^*$+C$^*$ $\rightarrow$ C$^*$+2H$_2$O(g), CO$_2^*$ $\rightarrow$ CO$_2$ (g))}
as depicted in {Figure 1} in supporting information (refer to this figure for a more detailed understanding, as it highlights all intermediates along with their structural views).

\begin{figure*}
\centering
{\includegraphics[trim=0.150cm 0.0cm 0.3cm 0.20cm,clip=true,width=1.00\textwidth]{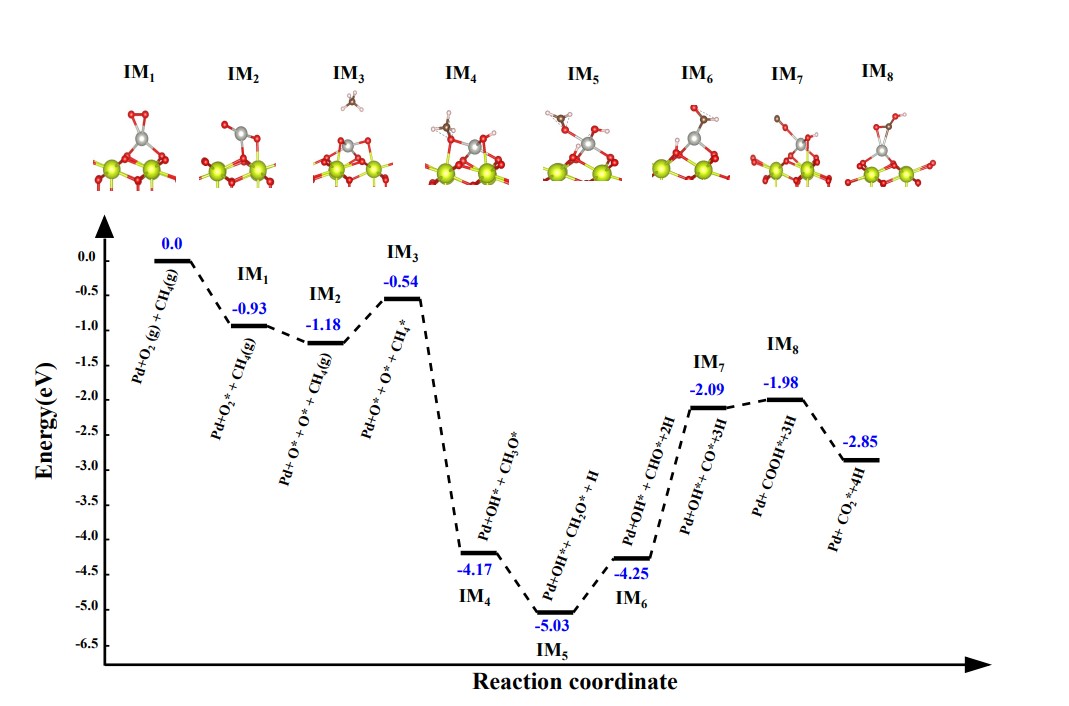}}\caption{The alternate reaction pathways of methane conversion into CO$_2$ on Pd1\text{@}CeO$_2$(111) in presence of oxygen with the structural view of corresponding intermediates.}
\label{Fig8}
\end{figure*}
\subsection{Alternate way for CO$_2$ Formation}
The complete reaction pathways for methane oxidation in oxygen-rich conditions were discussed in detail in an earlier section. Out of all the reaction steps, some of them are thermodynamically unfavorable, as represented in {Figure~\ref{Fig6}}.  Therefore, we tried to investigate the alternate path of CO$_2$ formation which shows more thermodynamical favorability. The alternate reaction path for CO$_2$ formation has been shown in {Figure~\ref{Fig8}}, which shows the conversion of CH$_4$ molecule in the presence of O$_2$ into CO$_2$ and H$_2$. The total Gibbs free energy during the complete oxidation of methane is -2.8 eV which is lower than the previous path (-6.3 eV). In this reaction pathway, after methane dissociates on the surface over two dissociated O$^*$ species acting as active sites, both CH$_3^*$ and H$^*$ bind to the dissociated O$^*$ species, forming CH$_3$O$^*$ and OH$^*$, respectively. Following the formation of OH$^*$, the next hydrogen atom is removed from CH$_3^*$ and bonds with the surface rather than the active site. The reaction proceeds with CH$_3$O$^*$ transforming stepwise into CH$_2$O$^*$, CHO$^*$, and eventually CO$^*$. After CO$^*$ and OH$^*$ are formed, they combine to form COOH$^*$, which, upon the removal of one hydrogen, produces CO$_2^*$.

\begin{figure*}
\hspace{-1.30 cm}
{\includegraphics[trim=0.15cm 0.0cm 0.0cm 0.0cm,clip=true,width=1.0\textwidth]{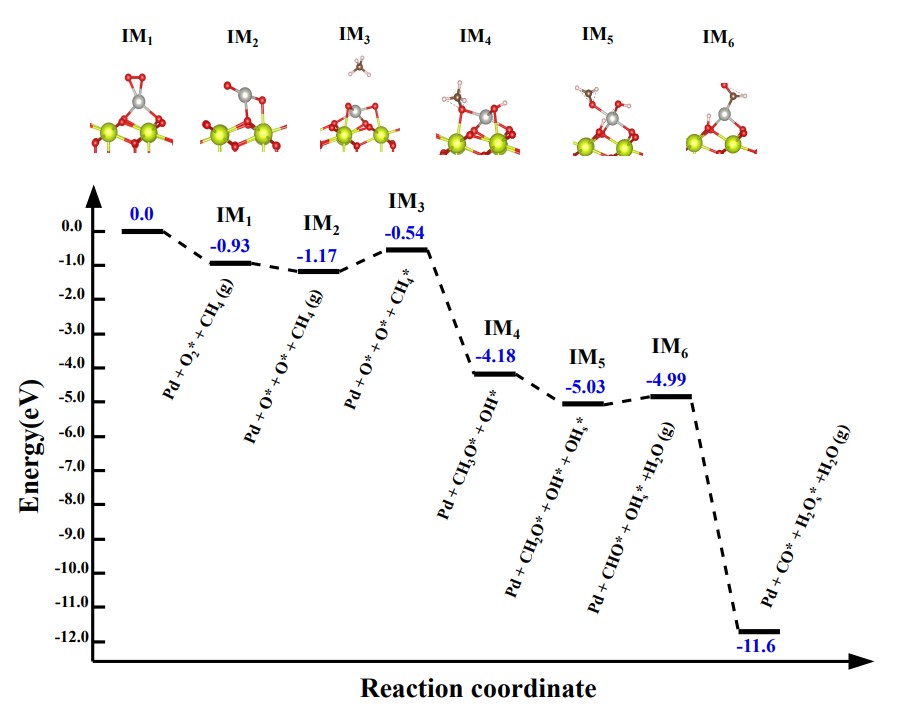}}
\caption{The role of lattice oxygen during methane oxidation. OH and OH$_s$ represent the formed species at the active site Pd and the surface, respectively.}
\label{Fig9}
\end{figure*}

\subsection{Role of Lattice Oxygen during CO Formation}
In our previous study, we reported the complete reaction path for methane oxidation without taking into account the lattice oxygen as an active site \cite{tomar}. Here, we investigated the role of lattice oxygen and discovered that, rather than leading to CO$_2$ formation, the lattice oxygen played a crucial role in the formation of CO. 
The formation of CO through methane oxidation is a thermodynamically favorable process with a reaction energy is -11.6 eV and the reaction pathway has been shown in {Figure~\ref{Fig9}}, where the dissociation steps of O$_2$ and CH$_4$ are identical to the previous reaction path ({Figure~\ref{Fig6}}). After CH$_4^*$ dissociates on O$^*$ species, the formation of two different CH$_3$O$^*$ and OH$^*$ occurs spontaneously. Later on, lattice oxygen plays a key role. CH$_3$O$^*$ loses one hydrogen, which bonds with the lattice oxygen, forming OH$_s^*$ (where 'OH$_s$' represents lattice oxygen). Similar to CH$_3$O$^*$, CH$_2$O$^*$, and CHO$^*$ remove hydrogen one by one and eventually CO$^*$ is formed.

\section{Microkinetic Modeling of Methane Oxidation on Pd1\text{@}CeO$_2$ and PdO1\text{@}CeO$_2$: Langmuir–Hinshelwood Mechanism and Reaction Pathway Analysis}
In the previous section, we extensively discussed various reaction pathways for complete methane oxidation on Pd1/PdO1\text{@}CeO$_2$(111) using DFT. Beyond DFT, we also employed microkinetic modeling to analyze the reaction dynamics during methane oxidation. This included examining the reaction rate, production rate, and degree of rate control, utilizing DFT-calculated parameters such as activation barriers, Gibbs free energy, and desorption energies. To provide clearer insight, we simulated the microkinetic model for Pd1\text{@}CeO$_2$ alone, as it features a single pathway with several intermediates that help illustrate the overall reaction. PdO1\text{@}CeO$_2$ follows a Langmuir–Hinshelwood mechanism during the methane oxidation process. Initially, both CH$_4$ and O$_2$ are adsorbed onto the surface. O$_2$ then dissociates into atomic O* species. CH$_4$ is subsequently decomposed into C$^*$ and H$^*$ species. The C$^*$ reacts with O$^*$ to produce CO$^*$, and CO$^*$ again reacts with O$^*$ and forms CO$_2$. The dissociated H$^*$ atoms combine with another O$^*$ to form H$_2$O. The desorption process is not detailed here, as H$_2$O and CO$_2$ are released directly into the gas phase. Given the minimal energy barrier for H migration on the Pd surface, it is reasonable to assume that the movement of H on the Pd surface has a negligible impact on the overall reaction cycle.

\begin{figure*}
\centering
{\includegraphics[trim=0.0cm 0.0cm 0.0cm 0.0cm,clip=true,width=0.80\textwidth]{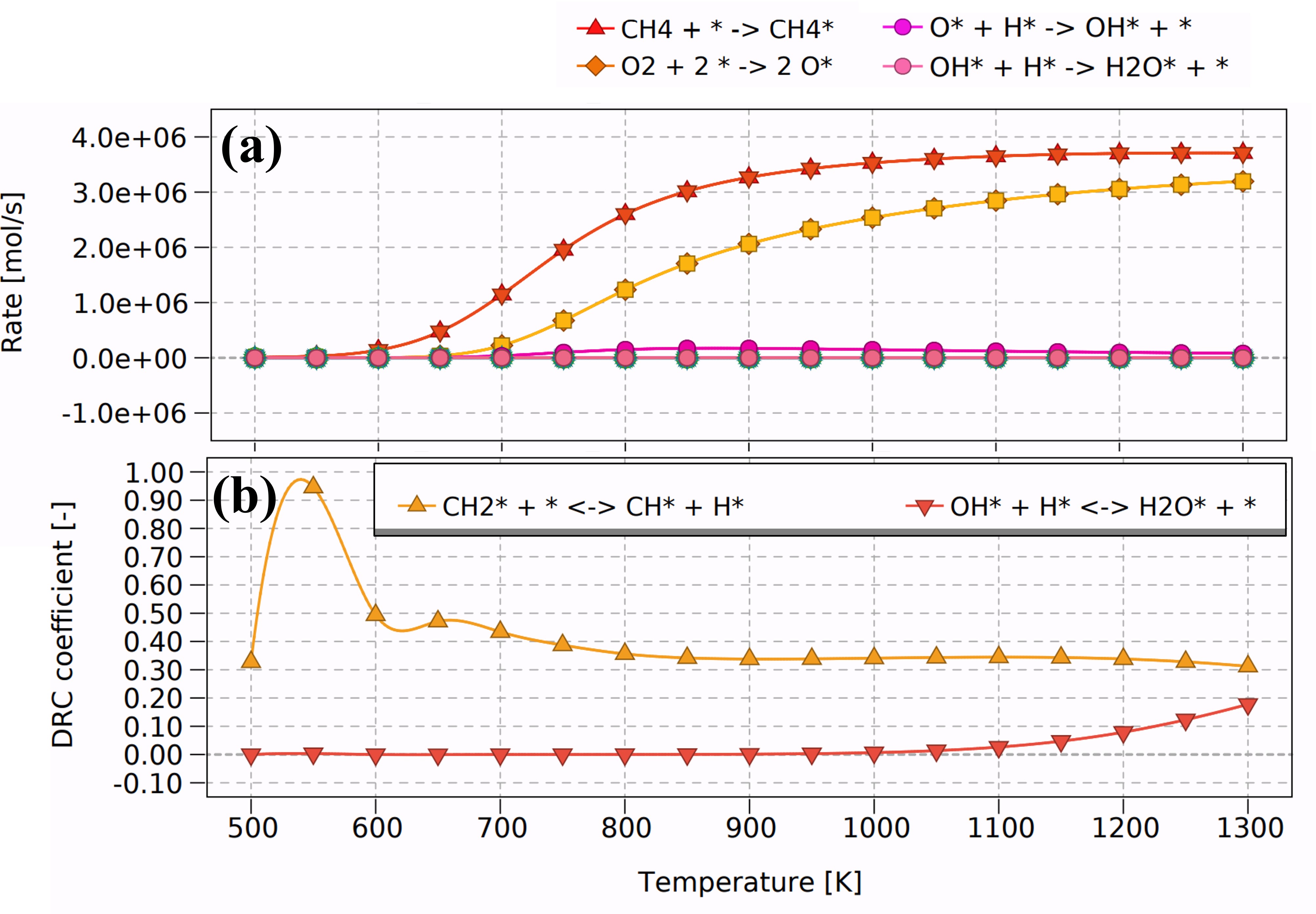}}
\caption{Reaction rate (a) and degree of rate control (b) as a function of temperature (in K) for methane conversion into CO$_2$ and H$_2$O on Pd1\text{@}CeO$_2$(111). These reaction steps play a significant role in controlling the overall reaction rate.}
\label{Fig10}
\end{figure*}
In {Figure~\ref{Fig10}(a) \& (b)}, the reaction rate and degree of rate control are plotted against temperature.
It is observed from the results that methane adsorption leads to much higher reaction rates than O$_2$ dissociation. The rate of methane adsorption increases exponentially between 600 and 1000 K, after which it levels off. In contrast, the rate of O$_2$ dissociation rises almost linearly from 700 K to 1300 K. The corresponding rates for associated reaction differ significantly, reflecting a notable variation in the composition of surface species required for the formation of CO$_2$ and H$_2$O. In the degree of rate control analysis, we focus on the key reaction steps that play a major role, specifically CH$_2^*$ $\rightarrow$ CH$^*$ + H$^*$ and OH$^*$ + H$^*$ $\rightarrow$ H$_2$O$^*$. Among these, the rate control for CH$_2$ dissociation is significant due to its high activation barrier, indicating that it largely governs the overall methane oxidation reaction. Additionally, the degree of rate control for H$_2$O$^*$ formation increases only at higher temperatures. 

\begin{figure*}
\centering
{\includegraphics[trim=0.150cm 0.20cm 0.10cm 0.20cm,clip=true,width=0.90\textwidth]{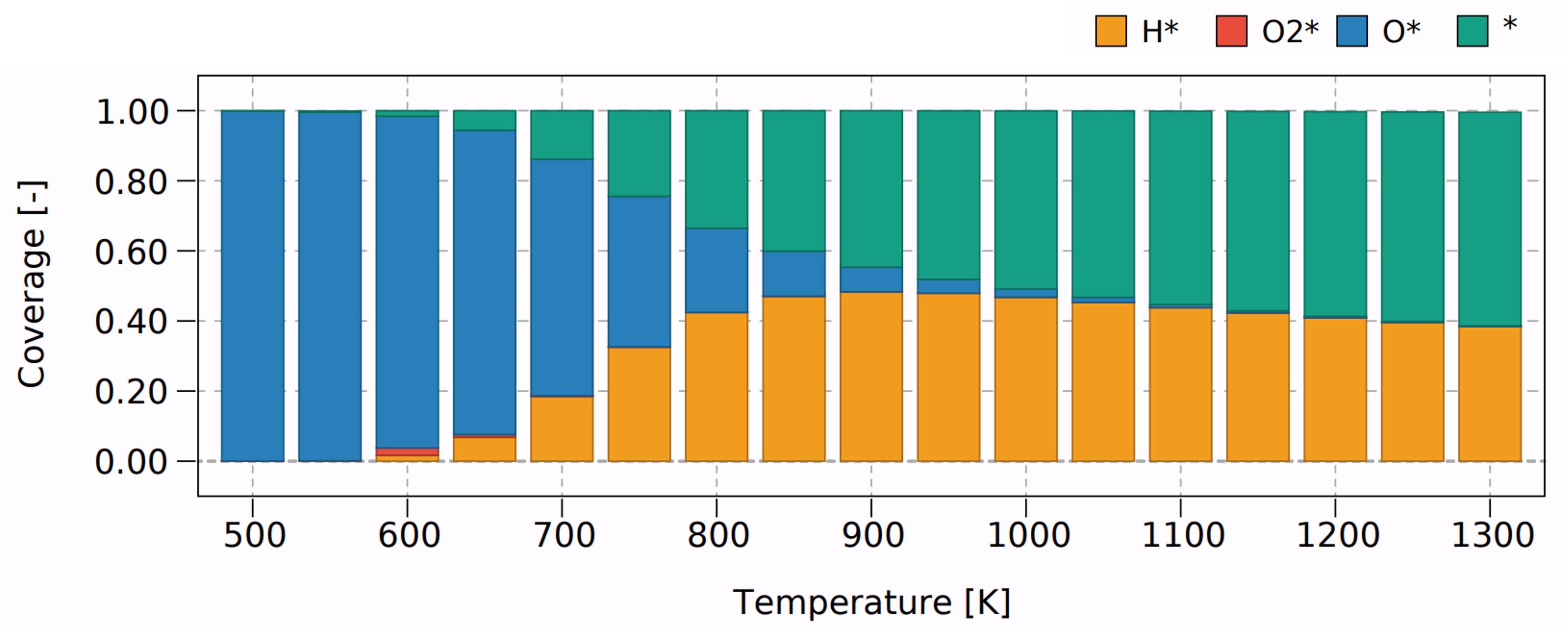}}
\caption{The surface coverage of intermediates as a function of temperature (in K) on Pd1\text{@}CeO$_2$(111)}
\label{Fig11}
\end{figure*}

As the rate of each elementary reaction step not only depends on the rate constant but also on the surface coverage of the involved intermediates and the microkinetic modeling provides insight into surface coverages, with the most abundant surface species shown in {Figure~\ref{Fig11}}. The surface is primarily covered by O$^*$ and H$^*$, with oxygen dominating at lower temperatures (around 500 K) and gradually decreasing as the temperature rises. In contrast, hydrogen coverage starts low and steadily increases, eventually covering nearly half of the surface at higher temperatures.

\begin{figure*}
\centering
{\includegraphics[trim=0.0cm 0.0cm 0.0cm 0.0cm,clip=true,width=0.90\textwidth]{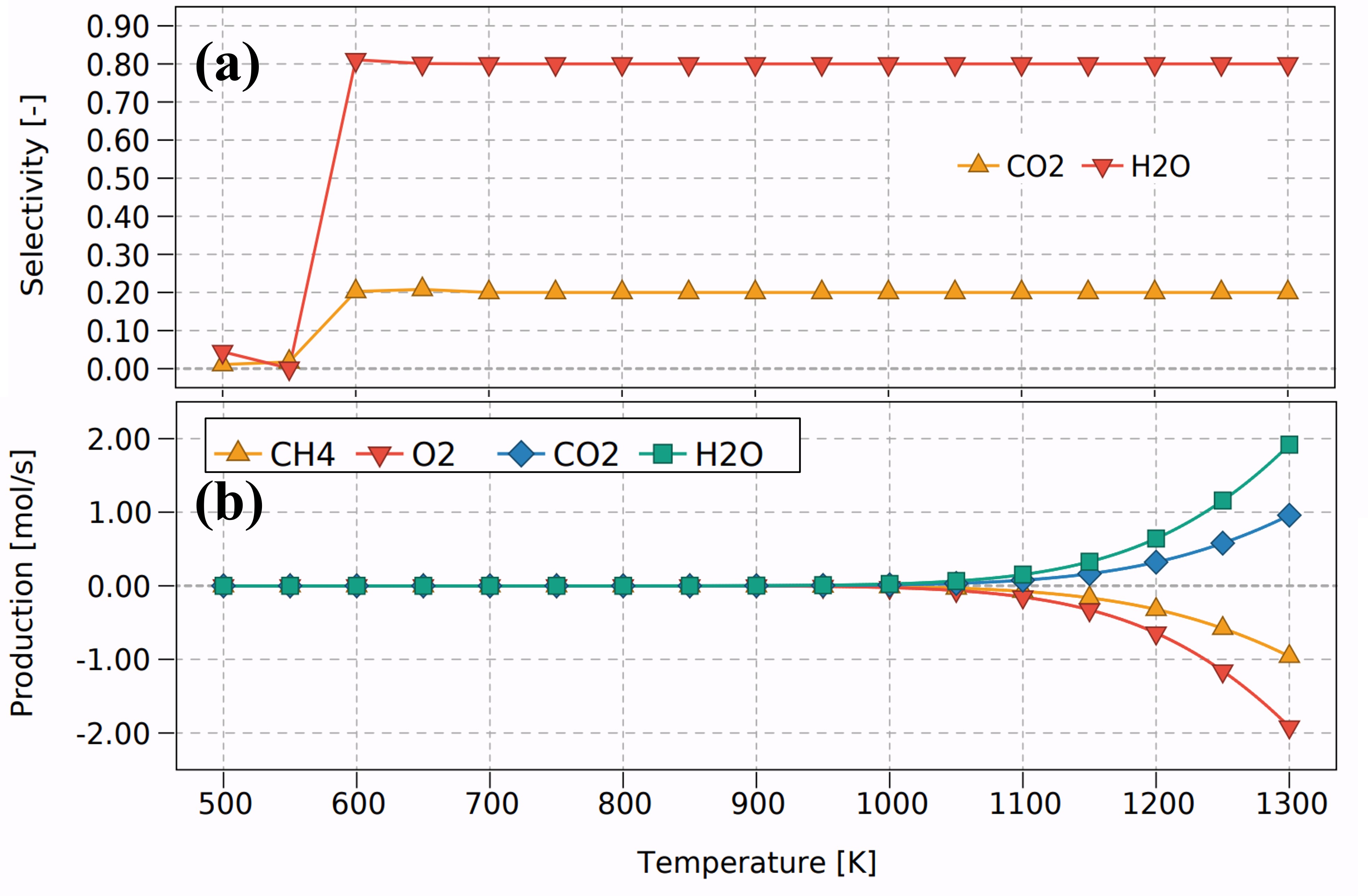}}
\caption{Microkinetic simulation for selectivity (a), production rate (b) as a function of the temperature between 500K and 1300K for products CO$_2$ and H$_2$O through the conversion of reactant CH$_4$ and O$_2$}
\label{Fig12}
\end{figure*}

In {Figure~\ref{Fig12}(a) \& (b)}, the selectivity and the production rates of CO$_2$ and H$_2$O are plotted concerning the methane oxidation process in the presence of O$_2$. The selectivity for H$_2$O formation is higher than CO$_2$ due to high H$^*$ surface coverage, and the selectivity of the product remains constant across different temperatures. At higher temperatures (around 1000 K), CH$_4$ and O$_2$ begin to decompose into CO$_2$ and H$_2$O. Notably, the decay rate of O$_2$ is faster than that of CH$_4$. 
Moreover, we investigated the apparent activation barrier as a function of temperature, as shown in {Figure~\ref{Fig13}}. The apparent barrier exhibits higher values around 500K to 600K, then decreases linearly beyond this range.

\begin{figure*}
\centering
{\includegraphics[trim=0.0cm 0.0cm 0.0cm 0.0cm,clip=true,width=0.90\textwidth]{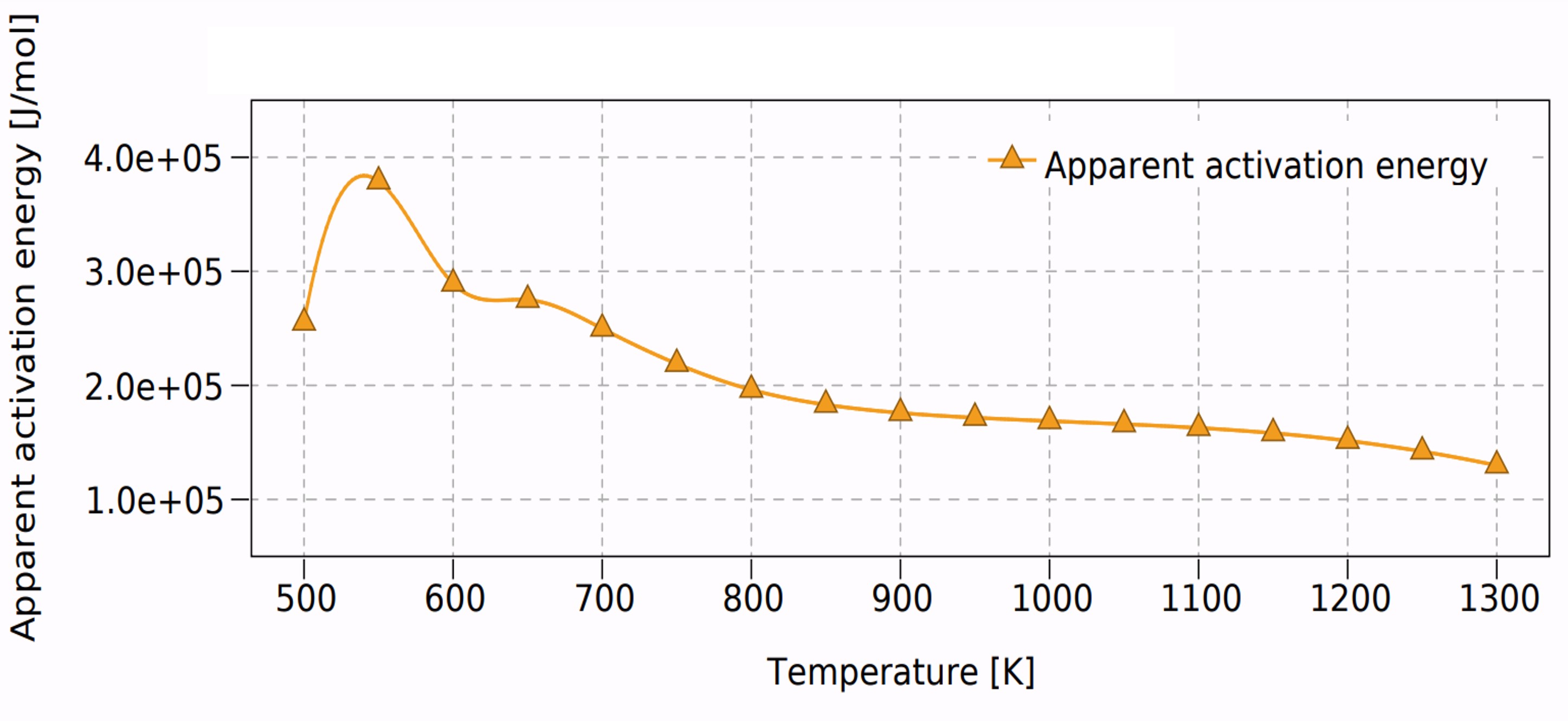}}
\caption{The apparent activation barrier as a function of the temperature (in K) for the methane oxidation into CO$_2$ and H$_2$O.}
\label{Fig13}
\end{figure*}

To compare the effect of binding between the C$^*$ and O$*$ species on the turnover frequency, we plotted the turnover frequency graph with respect to the binding energy of CO$^*$ and O$^*$, which is represented in {Figure~\ref{Fig14}}. It is worth to mentioning that turnover frequency in the presence of oxygen is higher, which indicates that the excess presence of oxygen helps to oxidation of methane into CO$_2$ and H$_2$O.

\begin{figure*}
\centering
{\includegraphics[trim=0.15cm 0.2cm 0.1cm 0.2cm,clip=true,width=0.60\textwidth]{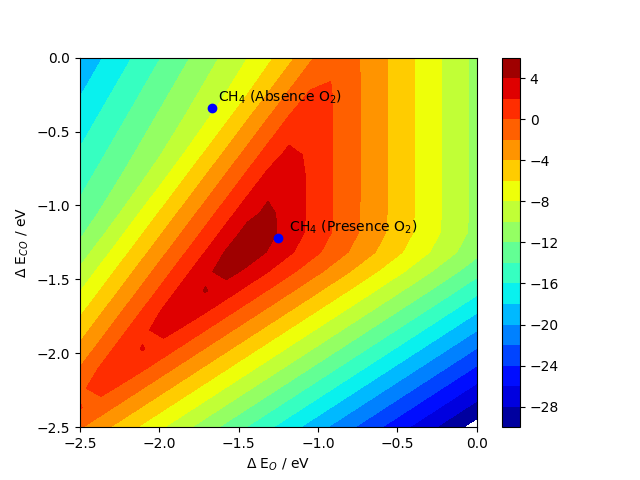}}
\caption{The turnover frequency of CO$_2$ formation concerning CO* and O* binding energy based on microkinetic modeling.}
\label{Fig14}
\end{figure*}
\clearpage
\newpage
\section{Machine learning of rate constants via symbolic regression using compressed sensing}

To model the rate constants for the reaction steps depicted in {Figure~\ref{Fig6}}, we employ the Sure Independence Screening and Sparsifying Operator (SISSO) method~\cite{ouyang2018sisso}, a machine learning approach based on symbolic regression via compressed sensing. This approach is particularly advantageous as it not only provides highly predictive models but also yields interpretable mathematical expressions for the descriptors involved. In the SISSO algorithm, we start with four primary features critical to the reaction kinetics:
\begin{itemize}
\item \(Q_1\): the charge on Pd, representing the electron density on the catalyst surface.
\item \(Q_2\): the charge on the reaction intermediate, indicating its interaction strength with the catalyst.
\item \(Z_N\): the coordination number of Pd, reflecting the local geometric environment of the active site.
\item \(d\): the distance between the reaction intermediate and Pd, capturing the spatial arrangement affecting orbital overlap.
\end{itemize}

\begin{table}
\centering
\begin{tabular}{|l|r|r|r|r|r|}
\hline
Reaction Intermediates &  d &      Q$_1$ &      Q$_2$ &  Z$_N$ &     log(K) \\
\hline
                  CH4* &    2.68000 & 9.751262 & 4.076353 &    2 &  37.631182 \\
                  CH3* &    2.00431 & 9.442062 & 4.280080 &    5 &  97.184808 \\
                  CH2* &    1.83909 & 9.368178 & 4.305077 &    3 & 152.453042 \\
                   CH* &    1.75211 & 9.332893 & 4.276711 &    3 & 192.540078 \\
                    C* &    1.64919 & 8.857009 & 4.214947 &    4 & 184.754929 \\
                    H* &    1.67875 & 9.378483 & 1.114597 &    3 & 131.042783 \\
                   O2* &    2.00594 & 9.256038 & 6.240521 &    4 &  45.804522 \\
                    O* &    1.80488 & 9.057392 & 6.571158 &    4 &  49.294294 \\
                   OH* &    1.98100 & 9.429129 & 7.479224 &    3 & 109.637353 \\
                  H2O* &    2.23677 & 9.484897 & 8.007955 &    3 & 111.191914 \\
                   CO* &    1.88227 & 9.686869 & 2.415333 &    3 &  83.170695 \\
\hline
\end{tabular}
\caption{Features and target (log(K)) used for the SISSO study.}
\label{tab:sisso_features}
\end{table}

These descriptors capture the essential electronic and geometric properties influencing the rate constants~\cite{bhattacharjee2016nh3}. The feature space is generated by systematically applying a set of mathematical operators \(\mathcal{H}^{(m)} = \{+, -, \times, \div, \sqrt{\phantom{x}}, (\cdot)^2, (\cdot)^3, \exp, \sqrt[3]~\}\) to the primary features, considering up to four iterations. This means that operators are applied recursively up to four times, creating new features from existing ones and expanding the feature space to capture higher-order and nonlinear relationships between the descriptors and the target property.

After four iterations, the recursive application of the operators to the four primary features resulted in a large feature space containing \(M\) candidate features. The sensing matrix \(\mathbf{D} \in \mathbf{R}^{N \times M}\) is constructed by evaluating these features for each of the \(N\) data points in our dataset. Each column of \(\mathbf{D}\) corresponds to a specific derived feature, and each row represents a data point with its computed feature values.

The optimization problem solved by SISSO is formulated as~\cite{ouyang2019simultaneous}:

\begin{equation}
\underset{\mathbf{c}}{\operatorname{argmin}} \left( || \mathbf{P} - \mathbf{D} \mathbf{c} ||_2^2 + \lambda || \mathbf{c} ||_0 \right),
\end{equation}

where \(\mathbf{P} = [\log(K_1), \log(K_2), \dots, \log(K_N)]\) is the target property vector consisting of the logarithms of the rate constants obtained via micro-kinetic modeling for the shown in the {Figure 6}. \(\mathbf{c} \in \mathbf{R}^M\) is the coefficient vector. \(\lambda\) is the regularization parameter enforcing sparsity. The \(l_0\) norm \(|| \mathbf{c} ||_0\) counts the number of non-zero components in \(\mathbf{c}\), promoting a compact and interpretable model.

Limiting the feature construction to four iterations strikes a balance between capturing complex relationships and maintaining model interpretability. While more iterations could generate additional features, they may lead to overfitting and reduce the transparency of the model.

Using the SISSO approach, we successfully generated a 4D descriptor for the \(\log(K)\) values, yielding a small error. The final expression is given by:

\begin{equation}
\log(K) = -0.7551 \left(\frac{1}{Q_1 \log(d)}\right) - \frac{3012.5 \cdot Q_2}{Z_N \exp(Q_2)} + 49.91 \left(\frac{Q_2^2}{\exp(Z_N)}\right) + 3726.38 \left(\frac{Q_1 - Q_2}{d^3}\right),
\label{4d}
\end{equation}

Each term in the expression for \(\log(K)\) was derived from combinations of the primary features through iterative application of the operators:

The first term \(-0.7551 \left(\frac{1}{Q_1 \log(d)}\right)\) captures a nonlinear relationship between the charge on Pd and the geometric factor \(d\). While the second term \(- \frac{3012.5 \cdot Q_2}{Z_N \exp(Q_2)}\) involves division and an exponential function, likely arising from higher-order iterations. It highlights the suppressive effect of strong interactions with the intermediate, modulated by the coordination number.
The third term \(+ 49.91 \left(\frac{Q_2^2}{\exp(Z_N)}\right)\) includes a squared term and an exponential, indicating a complex relationship derived over multiple iterations. It emphasizes how the charge on the intermediate and the Pd coordination influence \(K\).
The fourth term  \(+ 3726.38 \left(\frac{Q_1 - Q_2}{d^3}\right)\) combines subtraction, division, and exponentiation, likely from the fourth iteration. It represents the interplay between electronic differences and spatial factors.

Each primary feature in the above expression plays a significant role in influencing the rate constant $K$, reflecting underlying chemical interactions. For example, Increasing $Q1$ raises $log(K)$, leading to an increase in $K$. A higher electron density on Pd enhances catalytic activity by facilitating electron transfer processes. Specifically, in the first term, a larger $Q_1$ reduces the magnitude of the negative contribution $\left( \frac{1}{Q_1 \log(d)} \right)$, thus increasing $log(K)$. Similarly, in the fourth term, a higher $Q_1$ contributes positively to $\log(K)$. Similarly, one can see that increasing $Q_2$ has a dual effect. In the second term, a larger $Q_2$ increases the magnitude of the negative contribution due to the exponential term $\exp(Q_2)$, potentially decreasing $\log(K)$. However, the third term contains $Q_2^2$ in the numerator, and increasing $Q_2$ enhances this positive contribution, which can outweigh the negative effects. Overall, higher $Q_2$ tends to increase $K$ through its positive squared term, indicating that certain charge states of the intermediate favor reaction progression. It is easily understandable that a shorter distance enhances the overlap of electronic orbitals between the reaction intermediates and the catalyst surface, facilitating the reaction. In the first term, a larger $d$ increases $\log(d)$, which increases the denominator and increases the magnitude of the negative contribution. In the fourth term, a larger $d$ increases $d^3$ in the denominator, reducing the positive contribution to $\log(K)$.

From the above, it can be understood that optimizing these variables can significantly increase the reaction rate constant. Understanding how each variable influences $K$ allows for targeted modifications to the catalyst or reaction conditions to enhance performance. An interesting aspect of this model is the balance between the second and fourth terms, which have opposing signs and large coefficients. The second term has a strong negative influence on \(\log(K)\), particularly when \(Q_2\) is large, reflecting the suppressive effect of strong intermediate interactions. In contrast, the fourth term has a positive contribution, enhancing the rate constant when the difference \(Q_1 - Q_2\) is large or the distance \(d\) is small. This balance highlights the interplay between the electronic properties of Pd and the reaction intermediate, along with geometric factors. The performance of the 4D model is summarized in {Figure~\ref{SISSO}}, where the predicted \(\log(K)\) values are plotted against the reference microkinetic values. The model shows excellent agreement, with an \(R^2\) value of 0.997 and a Root Mean Squared Error (RMSE) of 2.851.

\begin{figure}
        \centering
    \includegraphics[width=0.85\linewidth]{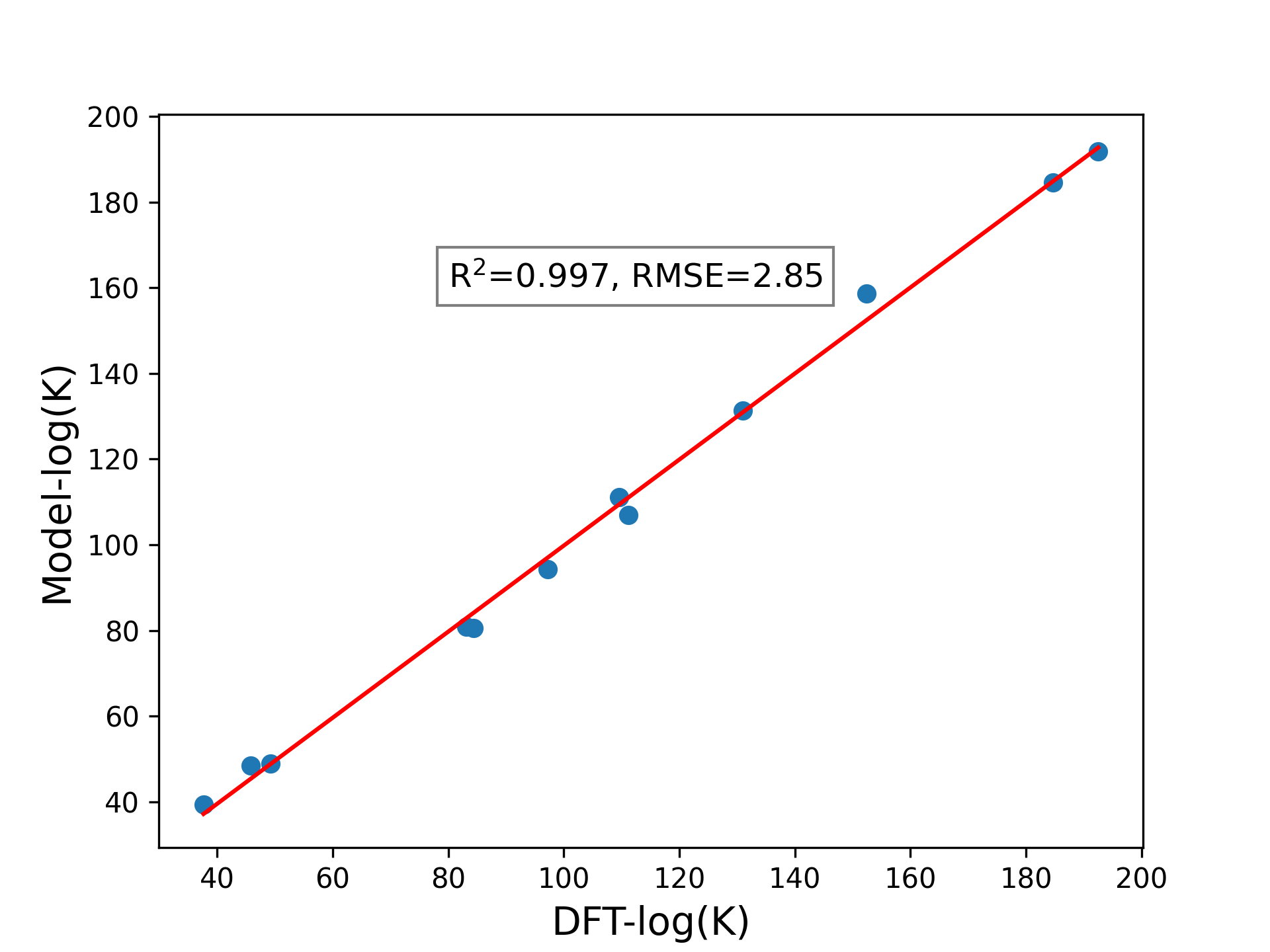}
    \caption{SISSO model performance for the 4D descriptor shown in Eq.~\ref{4d}. The predicted \(\log(K)\) values closely match the reference values, demonstrating the model's accuracy.}
    \label{SISSO}
\end{figure}
\clearpage
\newpage
\subsection*{Direction for the development of new catalysts via insights from the ML model}
The four-dimensional descriptor obtained from our SISSO model, comprising the Pd atomic charge ($Q_1$), the charge on the key CH$_x$O* intermediate ($Q_2$), the local Pd coordination number (ZN), and the Pd--intermediate distance ($d$)---offers physically interpretable guidance for rational catalyst design. Specifically, the algebraic form of Eq.~\ref{4d} reveals that high catalytic activity is associated with electron-rich Pd centers that remain in close proximity to moderately charged intermediates. This insight motivates three concrete design strategies. First, the term involving $(Q_1 - Q_2)/d^3$ suggests that activity can be enhanced by maximizing the electronic contrast between Pd and the adsorbate while minimizing the bonding distance. This can be achieved by introducing mild electron-donating dopants into the CeO$_2$ support or by alloying Pd with more electronegative elements (e.g., Pt, Rh) that enhance $Q_1$ without significantly altering $d$. Second, the exponential dependence on ZN in the model indicates that a moderate Pd coordination number (ZN $\approx$ 3--4) is optimal. Extremely undercoordinated Pd atoms are penalized due to destabilizing effects on transition states, while highly coordinated atoms suppress the favorable interaction terms. This coordination environment is naturally achieved in sub-nanometer Pd clusters or step-edge sites on CeO$_2$(111). Third, since overly polarized intermediates with large $Q_2$ values reduce the rate through the strongly negative exponential term, strategies that stabilize charge-neutral transition states---such as the engineering of oxygen vacancies or doping in the support to control the charge redistribution---can keep $Q_2$ within the favorable range of 1--3 $e$. 

Altogether, the above findings suggest that catalysts incorporating bimetallic PdM (M = Pt, Rh, Cu) clusters supported on vacancy-rich or doped CeO$_2$ surfaces are promising candidates for optimizing the key descriptors $(Q_1, Q_2, \mathrm{ZN}, d)$ identified by our model. This provides a systematic pathway towards the identification of the catalysts with improved performance for methane oxidation, grounded in interpretable machine learning and electronic structure insights.
\section{Conclusions}
In conclusion, this study delved into the catalytic mechanism of methane oxidation in the presence of oxygen through two distinct approaches. The first approach involved considering the oxygen-containing molecule PdO, while the second approach focused on the interaction of oxygen as a molecule with single-atom Pd supported with CeO$_2$(111). Utilizing detailed DFT calculations, we unraveled the complete pathway of methane dissociation into CO$_2$ under oxygen-rich conditions over 
 Pd1\text{@}CeO$_2$(111). Our results indicate that PdO exhibits favorable thermodynamics for the methane oxidation process over CeO$_2$(111). This study also highlights the crucial role played by a single Pd atom as an active site for oxygen dissociation, ultimately enhancing catalytic efficiency by reducing the activation barrier for methane oxidation by 0.14 eV. Our findings provide valuable insights into the fundamental processes involved in methane oxidation, which have implications for environmental remediation and industrial catalysis. 
In addition to the DFT analysis, we examined the apparent activation barrier, degree of rate control, and product selectivity using the microkinetic model. Our results show that surface coverage of H$^*$ increases with temperature, which enhances the production of H$_2$O, as confirmed by the product selectivity. The production CO$_2$ and H$_2$ also increase concerning temperature.

\begin{acknowledgement}
This work was supported financially by the Fundamental Research Program (PNK9750) of the Korea Institute of Materials Science (KIMS), South Korea. The authors also would like to acknowledge the Param Sanganak, IIT Kanpur supercomputing facility under the National Supercomputing Scheme (CDAC-NSM) for computing time.    
\end{acknowledgement}

\section*{Conflicts of interest}
The authors declare no conflict of interest.

\begin{suppinfo}
The supporting information included details of the value of adsorption energy and crystal structural view for different intermediate species on Pd and PdO supported with CeO$_2$(111) surface in {Tab: 1} and {Figure 1}, {Figure 2} \& {Figure 3}, respectively.

\end{suppinfo}

\bibliography{manuscript}

\providecommand{\latin}[1]{#1}
\makeatletter
\providecommand{\doi}
  {\begingroup\let\do\@makeother\dospecials
  \catcode`\{=1 \catcode`\}=2 \doi@aux}
\providecommand{\doi@aux}[1]{\endgroup\texttt{#1}}
\makeatother
\providecommand*\mcitethebibliography{\thebibliography}
\csname @ifundefined\endcsname{endmcitethebibliography}  {\let\endmcitethebibliography\endthebibliography}{}
\begin{mcitethebibliography}{89}
\providecommand*\natexlab[1]{#1}
\providecommand*\mciteSetBstSublistMode[1]{}
\providecommand*\mciteSetBstMaxWidthForm[2]{}
\providecommand*\mciteBstWouldAddEndPuncttrue
  {\def\EndOfBibitem{\unskip.}}
\providecommand*\mciteBstWouldAddEndPunctfalse
  {\let\EndOfBibitem\relax}
\providecommand*\mciteSetBstMidEndSepPunct[3]{}
\providecommand*\mciteSetBstSublistLabelBeginEnd[3]{}
\providecommand*\EndOfBibitem{}
\mciteSetBstSublistMode{f}
\mciteSetBstMaxWidthForm{subitem}{(\alph{mcitesubitemcount})}
\mciteSetBstSublistLabelBeginEnd
  {\mcitemaxwidthsubitemform\space}
  {\relax}
  {\relax}

\bibitem[Han \latin{et~al.}(2019)Han, Yang, and Han]{Zhan}
Han,~Z.; Yang,~Z.; Han,~M. Comprehensive investigation of methane conversion over Ni(111) surface under a consistent DFT framework: Implications for anti-coking of SOFC anodes. \emph{Applied Surface Science} \textbf{2019}, \emph{480}, 243--255\relax
\mciteBstWouldAddEndPuncttrue
\mciteSetBstMidEndSepPunct{\mcitedefaultmidpunct}
{\mcitedefaultendpunct}{\mcitedefaultseppunct}\relax
\EndOfBibitem
\bibitem[Lyubovsky and Pfefferle(1998)Lyubovsky, and Pfefferle]{Maxim}
Lyubovsky,~M.; Pfefferle,~L.~D. Methane combustion over the $\alpha$-alumina supported Pd catalyst: Activity of the mixed Pd/PdO state. \emph{Applied Catalysis A-general} \textbf{1998}, \emph{173}, 107--119\relax
\mciteBstWouldAddEndPuncttrue
\mciteSetBstMidEndSepPunct{\mcitedefaultmidpunct}
{\mcitedefaultendpunct}{\mcitedefaultseppunct}\relax
\EndOfBibitem
\bibitem[Rouet-Leduc and Hulbert(2024)Rouet-Leduc, and Hulbert]{Rouet}
Rouet-Leduc,~B.; Hulbert,~C. Automatic detection of methane emissions in multispectral satellite imagery using a vision transformer. \emph{Nature Communications} \textbf{2024}, \emph{15}, 3801\relax
\mciteBstWouldAddEndPuncttrue
\mciteSetBstMidEndSepPunct{\mcitedefaultmidpunct}
{\mcitedefaultendpunct}{\mcitedefaultseppunct}\relax
\EndOfBibitem
\bibitem[Mao \latin{et~al.}(2022)Mao, Zhang, Zhuang, Li, Liu, Zhou, Wang, Li, Lu, Liu, Montgomery, Joye, Zhang, and Yang]{Mao}
Mao,~S.-H.; Zhang,~H.-H.; Zhuang,~G.-C.; Li,~X.-J.; Liu,~Q.; Zhou,~Z.; Wang,~W.-L.; Li,~C.-Y.; Lu,~K.-Y.; Liu,~X.-T.; Montgomery,~A.; Joye,~S.~B.; Zhang,~Y.-Z.; Yang,~G.-P. Aerobic oxidation of methane significantly reduces global diffusive methane emissions from shallow marine waters. \emph{Nature Communications} \textbf{2022}, \emph{13}, 7309\relax
\mciteBstWouldAddEndPuncttrue
\mciteSetBstMidEndSepPunct{\mcitedefaultmidpunct}
{\mcitedefaultendpunct}{\mcitedefaultseppunct}\relax
\EndOfBibitem
\bibitem[{Le Mer} and Roger(2001){Le Mer}, and Roger]{jean}
{Le Mer},~J.; Roger,~P. Production, oxidation, emission and consumption of methane by soils: A review. \emph{European Journal of Soil Biology} \textbf{2001}, \emph{37}, 25--50\relax
\mciteBstWouldAddEndPuncttrue
\mciteSetBstMidEndSepPunct{\mcitedefaultmidpunct}
{\mcitedefaultendpunct}{\mcitedefaultseppunct}\relax
\EndOfBibitem
\bibitem[Liu \latin{et~al.}(2011)Liu, Zhang, Yan, Wang, and Xie]{Hongyan}
Liu,~H.; Zhang,~R.; Yan,~R.; Wang,~B.; Xie,~K. CH4 dissociation on NiCo (111) surface: A first-principles study. \emph{Applied Surface Science} \textbf{2011}, \emph{257}, 8955--8964\relax
\mciteBstWouldAddEndPuncttrue
\mciteSetBstMidEndSepPunct{\mcitedefaultmidpunct}
{\mcitedefaultendpunct}{\mcitedefaultseppunct}\relax
\EndOfBibitem
\bibitem[Burch \latin{et~al.}(1995)Burch, Urbano, and Loader]{Burch}
Burch,~R.; Urbano,~F.; Loader,~P. Methane combustion over palladium catalysts: The effect of carbon dioxide and water on activity. \emph{Applied Catalysis A: General} \textbf{1995}, \emph{123}, 173--184\relax
\mciteBstWouldAddEndPuncttrue
\mciteSetBstMidEndSepPunct{\mcitedefaultmidpunct}
{\mcitedefaultendpunct}{\mcitedefaultseppunct}\relax
\EndOfBibitem
\bibitem[Latimer \latin{et~al.}(2017)Latimer, Kulkarni, Aljama, Montoya, Yoo, Tsai, Abild-Pedersen, Studt, and Nørskov]{norskov}
Latimer,~A.~A.; Kulkarni,~A.~R.; Aljama,~H.; Montoya,~J.~H.; Yoo,~J.~S.; Tsai,~C.; Abild-Pedersen,~F.; Studt,~F.; Nørskov,~J.~K. Understanding trends in C–H bond activation in heterogeneous catalysis. \emph{Nature Materials} \textbf{2017}, \emph{16}, 225--229\relax
\mciteBstWouldAddEndPuncttrue
\mciteSetBstMidEndSepPunct{\mcitedefaultmidpunct}
{\mcitedefaultendpunct}{\mcitedefaultseppunct}\relax
\EndOfBibitem
\bibitem[Havran \latin{et~al.}(2011)Havran, Duduković, and Lo]{havran}
Havran,~V.; Duduković,~M.~P.; Lo,~C.~S. Conversion of Methane and Carbon Dioxide to Higher Value Products. \emph{Industrial \& Engineering Chemistry Research} \textbf{2011}, \emph{50}, 7089--7100\relax
\mciteBstWouldAddEndPuncttrue
\mciteSetBstMidEndSepPunct{\mcitedefaultmidpunct}
{\mcitedefaultendpunct}{\mcitedefaultseppunct}\relax
\EndOfBibitem
\bibitem[Webley and Tester(1991)Webley, and Tester]{webley}
Webley,~P.~A.; Tester,~J.~W. Fundamental kinetics of methane oxidation in supercritical water. \emph{Energy \& Fuels} \textbf{1991}, \emph{5}, 411--419\relax
\mciteBstWouldAddEndPuncttrue
\mciteSetBstMidEndSepPunct{\mcitedefaultmidpunct}
{\mcitedefaultendpunct}{\mcitedefaultseppunct}\relax
\EndOfBibitem
\bibitem[Yin \latin{et~al.}(2023)Yin, Pu, Xue, Ma, Wu, Han, Lin, Luo, Zeng, Ma, and Li]{Yin}
Yin,~H.; Pu,~Z.; Xue,~J.; Ma,~P.; Wu,~B.; Han,~M.; Lin,~H.; Luo,~Z.; Zeng,~J.; Ma,~X.; Li,~H. Oxygen Vacancy-Rich TiO2 as an Efficient Non-noble Metal Catalyst toward Mild Oxidation of Methane Using Hydrogen Peroxide as the Oxidant. \emph{ACS Catalysis} \textbf{2023}, \emph{13}, 7608--7615\relax
\mciteBstWouldAddEndPuncttrue
\mciteSetBstMidEndSepPunct{\mcitedefaultmidpunct}
{\mcitedefaultendpunct}{\mcitedefaultseppunct}\relax
\EndOfBibitem
\bibitem[Li \latin{et~al.}(2020)Li, Wang, Roy, van Bokhoven, and Artiglia]{xiansheng}
Li,~X.; Wang,~X.; Roy,~K.; van Bokhoven,~J.~A.; Artiglia,~L. Role of Water on the Structure of Palladium for Complete Oxidation of Methane. \emph{ACS Catalysis} \textbf{2020}, \emph{10}, 5783--5792\relax
\mciteBstWouldAddEndPuncttrue
\mciteSetBstMidEndSepPunct{\mcitedefaultmidpunct}
{\mcitedefaultendpunct}{\mcitedefaultseppunct}\relax
\EndOfBibitem
\bibitem[Li \latin{et~al.}(2021)Li, Beck, Krumeich, Artiglia, Ghosalya, Roger, Ferri, Kröcher, Sushkevich, Safonova, and van Bokhoven]{teng}
Li,~T.; Beck,~A.; Krumeich,~F.; Artiglia,~L.; Ghosalya,~M.~K.; Roger,~M.; Ferri,~D.; Kröcher,~O.; Sushkevich,~V.; Safonova,~O.~V.; van Bokhoven,~J.~A. Stable Palladium Oxide Clusters Encapsulated in Silicalite-1 for Complete Methane Oxidation. \emph{ACS Catalysis} \textbf{2021}, \emph{11}, 7371--7382\relax
\mciteBstWouldAddEndPuncttrue
\mciteSetBstMidEndSepPunct{\mcitedefaultmidpunct}
{\mcitedefaultendpunct}{\mcitedefaultseppunct}\relax
\EndOfBibitem
\bibitem[Zhu \latin{et~al.}(2013)Zhu, Zhang, Shan, Nguyen, Zhan, Gu, and Tao]{zhu}
Zhu,~Y.; Zhang,~S.; Shan,~J.-j.; Nguyen,~L.; Zhan,~S.; Gu,~X.; Tao,~F.~F. In Situ Surface Chemistries and Catalytic Performances of Ceria Doped with Palladium, Platinum, and Rhodium in Methane Partial Oxidation for the Production of Syngas. \emph{ACS Catalysis} \textbf{2013}, \emph{3}, 2627--2639\relax
\mciteBstWouldAddEndPuncttrue
\mciteSetBstMidEndSepPunct{\mcitedefaultmidpunct}
{\mcitedefaultendpunct}{\mcitedefaultseppunct}\relax
\EndOfBibitem
\bibitem[Willis \latin{et~al.}(2017)Willis, Gallo, Sokaras, Aljama, Nowak, Goodman, Wu, Tassone, Jaramillo, Abild-Pedersen, and Cargnello]{willis}
Willis,~J.~J.; Gallo,~A.; Sokaras,~D.; Aljama,~H.; Nowak,~S.~H.; Goodman,~E.~D.; Wu,~L.; Tassone,~C.~J.; Jaramillo,~T.~F.; Abild-Pedersen,~F.; Cargnello,~M. Systematic Structure–Property Relationship Studies in Palladium-Catalyzed Methane Complete Combustion. \emph{ACS Catalysis} \textbf{2017}, \emph{7}, 7810--7821\relax
\mciteBstWouldAddEndPuncttrue
\mciteSetBstMidEndSepPunct{\mcitedefaultmidpunct}
{\mcitedefaultendpunct}{\mcitedefaultseppunct}\relax
\EndOfBibitem
\bibitem[Kinnunen \latin{et~al.}(2011)Kinnunen, Hirvi, Suvanto, and Pakkanen]{kinnunen2011}
Kinnunen,~N.~M.; Hirvi,~J.~T.; Suvanto,~M.; Pakkanen,~T.~A. Role of the Interface between Pd and PdO in Methane Dissociation. \emph{The Journal of Physical Chemistry C} \textbf{2011}, \emph{115}, 19197--19202\relax
\mciteBstWouldAddEndPuncttrue
\mciteSetBstMidEndSepPunct{\mcitedefaultmidpunct}
{\mcitedefaultendpunct}{\mcitedefaultseppunct}\relax
\EndOfBibitem
\bibitem[Senftle \latin{et~al.}(2015)Senftle, van Duin, and Janik]{senftle2015}
Senftle,~T.~P.; van Duin,~A. C.~T.; Janik,~M.~J. Role of Site Stability in Methane Activation on PdxCe1xO Surfaces. \emph{ACS Catalysis} \textbf{2015}, \emph{5}, 6187--6199\relax
\mciteBstWouldAddEndPuncttrue
\mciteSetBstMidEndSepPunct{\mcitedefaultmidpunct}
{\mcitedefaultendpunct}{\mcitedefaultseppunct}\relax
\EndOfBibitem
\bibitem[Herron \latin{et~al.}(2012)Herron, Tonelli, and Mavrikakis]{HERRON2012}
Herron,~J.~A.; Tonelli,~S.; Mavrikakis,~M. Atomic and molecular adsorption on Pd(111). \emph{Surface Science} \textbf{2012}, \emph{606}, 1670--1679\relax
\mciteBstWouldAddEndPuncttrue
\mciteSetBstMidEndSepPunct{\mcitedefaultmidpunct}
{\mcitedefaultendpunct}{\mcitedefaultseppunct}\relax
\EndOfBibitem
\bibitem[Jørgensen and Gr{\"o}nbeck(2016)Jørgensen, and Gr{\"o}nbeck]{Mikkel2016}
Jørgensen,~M.; Gr{\"o}nbeck,~H. First-Principles Microkinetic Modeling of Methane Oxidation over Pd(100) and Pd(111). \emph{ACS Catalysis} \textbf{2016}, \emph{6}, 6730--6738\relax
\mciteBstWouldAddEndPuncttrue
\mciteSetBstMidEndSepPunct{\mcitedefaultmidpunct}
{\mcitedefaultendpunct}{\mcitedefaultseppunct}\relax
\EndOfBibitem
\bibitem[Kumar \latin{et~al.}(2022)Kumar, Al-Attas, Hu, and Kibria]{kumar}
Kumar,~P.; Al-Attas,~T.~A.; Hu,~J.; Kibria,~M.~G. Single Atom Catalysts for Selective Methane Oxidation to Oxygenates. \emph{ACS Nano} \textbf{2022}, \emph{16}, 8557--8618\relax
\mciteBstWouldAddEndPuncttrue
\mciteSetBstMidEndSepPunct{\mcitedefaultmidpunct}
{\mcitedefaultendpunct}{\mcitedefaultseppunct}\relax
\EndOfBibitem
\bibitem[Sun \latin{et~al.}(2024)Sun, Tu, Xu, Yang, Yu, Zhai, Ci, and Deng]{Sun}
Sun,~J.; Tu,~R.; Xu,~Y.; Yang,~H.; Yu,~T.; Zhai,~D.; Ci,~X.; Deng,~W. Machine learning aided design of single-atom alloy catalysts for methane cracking. \emph{Nature Communications} \textbf{2024}, \emph{15}, 6036\relax
\mciteBstWouldAddEndPuncttrue
\mciteSetBstMidEndSepPunct{\mcitedefaultmidpunct}
{\mcitedefaultendpunct}{\mcitedefaultseppunct}\relax
\EndOfBibitem
\bibitem[Feng \latin{et~al.}(2018)Feng, Wan, Xiong, Zhou, Chen, Pereira~Hernandez, Wang, Lin, Datye, and Guo]{Feng}
Feng,~Y.; Wan,~Q.; Xiong,~H.; Zhou,~S.; Chen,~X.; Pereira~Hernandez,~X.~I.; Wang,~Y.; Lin,~S.; Datye,~A.~K.; Guo,~H. Correlating DFT Calculations with CO Oxidation Reactivity on Ga-Doped Pt/CeO2 Single-Atom Catalysts. \emph{The Journal of Physical Chemistry C} \textbf{2018}, \emph{122}, 22460--22468\relax
\mciteBstWouldAddEndPuncttrue
\mciteSetBstMidEndSepPunct{\mcitedefaultmidpunct}
{\mcitedefaultendpunct}{\mcitedefaultseppunct}\relax
\EndOfBibitem
\bibitem[Wu \latin{et~al.}(2020)Wu, Hu, Yu, Shen, and Li]{Wu}
Wu,~L.; Hu,~S.; Yu,~W.; Shen,~S.; Li,~T. Stabilizing mechanism of single-atom catalysts on a defective carbon surface. \emph{npj Computational Materials} \textbf{2020}, \emph{6}, 23\relax
\mciteBstWouldAddEndPuncttrue
\mciteSetBstMidEndSepPunct{\mcitedefaultmidpunct}
{\mcitedefaultendpunct}{\mcitedefaultseppunct}\relax
\EndOfBibitem
\bibitem[Akri \latin{et~al.}(2019)Akri, Zhao, Li, Zang, Lee, Isaacs, Xi, Gangarajula, Luo, and et. al.]{Akri2019}
Akri,~M.; Zhao,~S.; Li,~X.; Zang,~K.; Lee,~A.~F.; Isaacs,~M.~A.; Xi,~W.; Gangarajula,~Y.; Luo,~J.; et. al. Atomically dispersed nickel as coke-resistant active sites for methane dry reforming. \emph{Nature Communications} \textbf{2019}, \emph{10}, 5181\relax
\mciteBstWouldAddEndPuncttrue
\mciteSetBstMidEndSepPunct{\mcitedefaultmidpunct}
{\mcitedefaultendpunct}{\mcitedefaultseppunct}\relax
\EndOfBibitem
\bibitem[Montini \latin{et~al.}(2016)Montini, Melchionna, Monai, and Fornasiero]{montini2016}
Montini,~T.; Melchionna,~M.; Monai,~M.; Fornasiero,~P. Fundamentals and Catalytic Applications of CeO2-Based Materials. \emph{Chemical Reviews} \textbf{2016}, \emph{116}, 5987--6041\relax
\mciteBstWouldAddEndPuncttrue
\mciteSetBstMidEndSepPunct{\mcitedefaultmidpunct}
{\mcitedefaultendpunct}{\mcitedefaultseppunct}\relax
\EndOfBibitem
\bibitem[Senftle \latin{et~al.}(2017)Senftle, Van~Duin, and Janik]{senftle2017methane}
Senftle,~T.~P.; Van~Duin,~A.~C.; Janik,~M.~J. Methane activation at the Pd/CeO2 interface. \emph{Acs Catalysis} \textbf{2017}, \emph{7}, 327--332\relax
\mciteBstWouldAddEndPuncttrue
\mciteSetBstMidEndSepPunct{\mcitedefaultmidpunct}
{\mcitedefaultendpunct}{\mcitedefaultseppunct}\relax
\EndOfBibitem
\bibitem[Lustemberg \latin{et~al.}(2016)Lustemberg, Ramírez, Liu, Guti{\'e}rrez, Grinter, Carrasco, Senanayake, Rodriguez, and Ganduglia-Pirovano]{pablo2016}
Lustemberg,~P.~G.; Ramírez,~P.~J.; Liu,~Z.; Guti{\'e}rrez,~R.~A.; Grinter,~D.~G.; Carrasco,~J.; Senanayake,~S.~D.; Rodriguez,~J.~A.; Ganduglia-Pirovano,~M.~V. Room-Temperature Activation of Methane and Dry Re-forming with CO2 on Ni-CeO2(111) Surfaces: Effect of Ce3+ Sites and Metal–Support Interactions on C–H Bond Cleavage. \emph{ACS Catalysis} \textbf{2016}, \emph{6}, 8184--8191\relax
\mciteBstWouldAddEndPuncttrue
\mciteSetBstMidEndSepPunct{\mcitedefaultmidpunct}
{\mcitedefaultendpunct}{\mcitedefaultseppunct}\relax
\EndOfBibitem
\bibitem[Lustemberg \latin{et~al.}(2021)Lustemberg, Mao, Salcedo, Irigoyen, Ganduglia-Pirovano, and Campbell]{pablo2021}
Lustemberg,~P.~G.; Mao,~Z.; Salcedo,~A.; Irigoyen,~B.; Ganduglia-Pirovano,~M.~V.; Campbell,~C.~T. Nature of the Active Sites on Ni/CeO2 Catalysts for Methane Conversions. \emph{ACS Catalysis} \textbf{2021}, \emph{11}, 10604--10613\relax
\mciteBstWouldAddEndPuncttrue
\mciteSetBstMidEndSepPunct{\mcitedefaultmidpunct}
{\mcitedefaultendpunct}{\mcitedefaultseppunct}\relax
\EndOfBibitem
\bibitem[Wang and Gong(2018)Wang, and Gong]{wang2018}
Wang,~J.; Gong,~X.-Q. A DFT+U study of V, Cr and Mn doped CeO2(111). \emph{Applied Surface Science} \textbf{2018}, \emph{428}, 377--384\relax
\mciteBstWouldAddEndPuncttrue
\mciteSetBstMidEndSepPunct{\mcitedefaultmidpunct}
{\mcitedefaultendpunct}{\mcitedefaultseppunct}\relax
\EndOfBibitem
\bibitem[Mao \latin{et~al.}(2017)Mao, {van Duin}, and Luo]{QianMao}
Mao,~Q.; {van Duin},~A.~C.; Luo,~K. Investigation of methane oxidation by palladium-based catalyst via ReaxFF Molecular Dynamics simulation. \emph{Proceedings of the Combustion Institute} \textbf{2017}, \emph{36}, 4339--4346\relax
\mciteBstWouldAddEndPuncttrue
\mciteSetBstMidEndSepPunct{\mcitedefaultmidpunct}
{\mcitedefaultendpunct}{\mcitedefaultseppunct}\relax
\EndOfBibitem
\bibitem[Hicks \latin{et~al.}(1990)Hicks, Qi, Young, and Lee]{RobertF}
Hicks,~R.~F.; Qi,~H.; Young,~M.~L.; Lee,~R.~G. Structure sensitivity of methane oxidation over platinum and palladium. \emph{Journal of Catalysis} \textbf{1990}, \emph{122}, 280--294\relax
\mciteBstWouldAddEndPuncttrue
\mciteSetBstMidEndSepPunct{\mcitedefaultmidpunct}
{\mcitedefaultendpunct}{\mcitedefaultseppunct}\relax
\EndOfBibitem
\bibitem[Shi \latin{et~al.}(2024)Shi, Han, Zang, Wang, Li, Zhang, and Liu]{shi}
Shi,~S.; Han,~Y.; Zang,~Y.; Wang,~Z.; Li,~Y.; Zhang,~H.; Liu,~Z. Identification of Active Phase for Complete Oxidation of Methane on Palladium Surface. \textbf{2024}, \emph{67}, 874--879\relax
\mciteBstWouldAddEndPuncttrue
\mciteSetBstMidEndSepPunct{\mcitedefaultmidpunct}
{\mcitedefaultendpunct}{\mcitedefaultseppunct}\relax
\EndOfBibitem
\bibitem[Datye \latin{et~al.}(2000)Datye, Bravo, Nelson, Atanasova, Lyubovsky, and Pfefferle]{Abhaya}
Datye,~A.~K.; Bravo,~J.; Nelson,~T.~R.; Atanasova,~P.; Lyubovsky,~M.; Pfefferle,~L. Catalyst microstructure and methane oxidation reactivity during the Pd-PdO transformation on alumina supports. \emph{Applied Catalysis A: General} \textbf{2000}, \emph{198}, 179--196\relax
\mciteBstWouldAddEndPuncttrue
\mciteSetBstMidEndSepPunct{\mcitedefaultmidpunct}
{\mcitedefaultendpunct}{\mcitedefaultseppunct}\relax
\EndOfBibitem
\bibitem[Chin \latin{et~al.}(2013)Chin, Buda, Neurock, and Iglesia]{Iglesia}
Chin,~Y.-H.~C.; Buda,~C.; Neurock,~M.; Iglesia,~E. Consequences of Metal–Oxide Interconversion for C–H Bond Activation during CH4 Reactions on Pd Catalysts. \emph{Journal of the American Chemical Society} \textbf{2013}, \emph{135}, 15425--15442\relax
\mciteBstWouldAddEndPuncttrue
\mciteSetBstMidEndSepPunct{\mcitedefaultmidpunct}
{\mcitedefaultendpunct}{\mcitedefaultseppunct}\relax
\EndOfBibitem
\bibitem[Chen \latin{et~al.}(2019)Chen, Lin, Chen, Chen, Xu, Wang, Zhang, and Zheng]{Chen}
Chen,~B.; Lin,~J.; Chen,~X.; Chen,~Y.; Xu,~Y.; Wang,~Z.; Zhang,~W.; Zheng,~Y. Cooperative Catalysis of Methane Oxidation through Modulating the Stabilization of PdO and Electronic Properties over Ti-Doped Alumina-Supported Palladium Catalysts. \emph{ACS Omega} \textbf{2019}, \emph{4}, 18582--18592\relax
\mciteBstWouldAddEndPuncttrue
\mciteSetBstMidEndSepPunct{\mcitedefaultmidpunct}
{\mcitedefaultendpunct}{\mcitedefaultseppunct}\relax
\EndOfBibitem
\bibitem[Wang and Wang(2022)Wang, and Wang]{wang1}
Wang,~J.; Wang,~G.-C. Dynamic Evolution of Methane Oxidation on Pd-Based Catalysts: A Reactive Force Field Molecular Dynamics Study. \emph{The Journal of Physical Chemistry C} \textbf{2022}, \emph{126}, 14201--14210\relax
\mciteBstWouldAddEndPuncttrue
\mciteSetBstMidEndSepPunct{\mcitedefaultmidpunct}
{\mcitedefaultendpunct}{\mcitedefaultseppunct}\relax
\EndOfBibitem
\bibitem[Du \latin{et~al.}(2020)Du, Zhao, Wang, Zhao, Li, and Luo]{Du}
Du,~J.; Zhao,~D.; Wang,~C.; Zhao,~Y.; Li,~H.; Luo,~Y. Size effects of Pd nanoparticles supported over CeZrPAl for methane oxidation. \emph{Catal. Sci. Technol.} \textbf{2020}, \emph{10}, 7875--7882\relax
\mciteBstWouldAddEndPuncttrue
\mciteSetBstMidEndSepPunct{\mcitedefaultmidpunct}
{\mcitedefaultendpunct}{\mcitedefaultseppunct}\relax
\EndOfBibitem
\bibitem[Petrov \latin{et~al.}(2018)Petrov, Ferri, Krumeich, Nachtegaal, van Bokhoven, and Kröcher]{Petrov}
Petrov,~A.~W.; Ferri,~D.; Krumeich,~F.; Nachtegaal,~M.; van Bokhoven,~J.~A.; Kröcher,~O. Stable complete methane oxidation over palladium based zeolite catalysts. \emph{Nature Communications} \textbf{2018}, \emph{9}, 2545\relax
\mciteBstWouldAddEndPuncttrue
\mciteSetBstMidEndSepPunct{\mcitedefaultmidpunct}
{\mcitedefaultendpunct}{\mcitedefaultseppunct}\relax
\EndOfBibitem
\bibitem[Zhang \latin{et~al.}(2024)Zhang, Zheng, Gao, Liu, Ji, Tian, Zou, Sun, Hu, Chen, Chen, Liu, Zhong, Xu, Zhu, and Su]{zhangten}
Zhang,~T. \latin{et~al.}  Simultaneously activating molecular oxygen and surface lattice oxygen on Pt/TiO2 for low-temperature CO oxidation. \emph{Nature Communications} \textbf{2024}, \emph{15}, 6827\relax
\mciteBstWouldAddEndPuncttrue
\mciteSetBstMidEndSepPunct{\mcitedefaultmidpunct}
{\mcitedefaultendpunct}{\mcitedefaultseppunct}\relax
\EndOfBibitem
\bibitem[Bunting \latin{et~al.}(2019)Bunting, Cheng, Thompson, and Hu]{Bunting}
Bunting,~R.~J.; Cheng,~X.; Thompson,~J.; Hu,~P. Amorphous Surface PdOX and Its Activity toward Methane Combustion. \emph{ACS Catalysis} \textbf{2019}, \emph{9}, 10317--10323\relax
\mciteBstWouldAddEndPuncttrue
\mciteSetBstMidEndSepPunct{\mcitedefaultmidpunct}
{\mcitedefaultendpunct}{\mcitedefaultseppunct}\relax
\EndOfBibitem
\bibitem[Xiong \latin{et~al.}(2021)Xiong, Kunwar, Jiang, García-Vargas, Li, Du, Canning, Pereira-Hernandez, and et. al.]{Xiong}
Xiong,~H.; Kunwar,~D.; Jiang,~D.; García-Vargas,~C.~E.; Li,~H.; Du,~C.; Canning,~G.; Pereira-Hernandez,~X.~I.; et. al. Engineering catalyst supports to stabilize PdOx two-dimensional rafts for water-tolerant methane oxidation. \emph{Nature Catalysis} \textbf{2021}, \emph{4}, 830--839\relax
\mciteBstWouldAddEndPuncttrue
\mciteSetBstMidEndSepPunct{\mcitedefaultmidpunct}
{\mcitedefaultendpunct}{\mcitedefaultseppunct}\relax
\EndOfBibitem
\bibitem[Yue \latin{et~al.}(2024)Yue, Praveen, Klyushin, Fedorov, Hashimoto, Li, Jones, Liu, Yu, Willinger, and Huang]{Yue}
Yue,~S.; Praveen,~C.~S.; Klyushin,~A.; Fedorov,~A.; Hashimoto,~M.; Li,~Q.; Jones,~T.; Liu,~P.; Yu,~W.; Willinger,~M.-G.; Huang,~X. Redox dynamics and surface structures of an active palladium catalyst during methane oxidation. \emph{Nature Communications} \textbf{2024}, \emph{15}, 4678\relax
\mciteBstWouldAddEndPuncttrue
\mciteSetBstMidEndSepPunct{\mcitedefaultmidpunct}
{\mcitedefaultendpunct}{\mcitedefaultseppunct}\relax
\EndOfBibitem
\bibitem[Hellman \latin{et~al.}(2012)Hellman, Resta, Martin, Gustafson, Trinchero, Carlsson, Balmes, Felici, van Rijn, Frenken, Andersen, Lundgren, and Grönbeck]{Hellman}
Hellman,~A.; Resta,~A.; Martin,~N.~M.; Gustafson,~J.; Trinchero,~A.; Carlsson,~P.-A.; Balmes,~O.; Felici,~R.; van Rijn,~R.; Frenken,~J. W.~M.; Andersen,~J.~N.; Lundgren,~E.; Grönbeck,~H. The Active Phase of Palladium during Methane Oxidation. \emph{The Journal of Physical Chemistry Letters} \textbf{2012}, \emph{3}, 678--682\relax
\mciteBstWouldAddEndPuncttrue
\mciteSetBstMidEndSepPunct{\mcitedefaultmidpunct}
{\mcitedefaultendpunct}{\mcitedefaultseppunct}\relax
\EndOfBibitem
\bibitem[Stotz \latin{et~al.}(2019)Stotz, Maier, Boubnov, Gremminger, Grunwaldt, and Deutschmann]{stotz}
Stotz,~H.; Maier,~L.; Boubnov,~A.; Gremminger,~A.; Grunwaldt,~J.-D.; Deutschmann,~O. Surface reaction kinetics of methane oxidation over PdO. \emph{Journal of Catalysis} \textbf{2019}, \emph{370}, 152--175\relax
\mciteBstWouldAddEndPuncttrue
\mciteSetBstMidEndSepPunct{\mcitedefaultmidpunct}
{\mcitedefaultendpunct}{\mcitedefaultseppunct}\relax
\EndOfBibitem
\bibitem[{Moncada Quintero} \latin{et~al.}(2021){Moncada Quintero}, Ercolino, Poozhikunnath, Maric, and Specchia]{carmen}
{Moncada Quintero},~C.~W.; Ercolino,~G.; Poozhikunnath,~A.; Maric,~R.; Specchia,~S. Analysis of heat and mass transfer limitations for the combustion of methane emissions on PdO/Co3O4 coated on ceramic open cell foams. \emph{Chemical Engineering Journal} \textbf{2021}, \emph{405}, 126970\relax
\mciteBstWouldAddEndPuncttrue
\mciteSetBstMidEndSepPunct{\mcitedefaultmidpunct}
{\mcitedefaultendpunct}{\mcitedefaultseppunct}\relax
\EndOfBibitem
\bibitem[Koo \latin{et~al.}(2017)Koo, Yu, Choi, Jang, Cheong, and Kim]{Koo}
Koo,~W.-T.; Yu,~S.; Choi,~S.-J.; Jang,~J.-S.; Cheong,~J.~Y.; Kim,~I.-D. Nanoscale PdO Catalyst Functionalized Co3O4 Hollow Nanocages Using MOF Templates for Selective Detection of Acetone Molecules in Exhaled Breath. \emph{ACS Applied Materials \& Interfaces} \textbf{2017}, \emph{9}, 8201--8210\relax
\mciteBstWouldAddEndPuncttrue
\mciteSetBstMidEndSepPunct{\mcitedefaultmidpunct}
{\mcitedefaultendpunct}{\mcitedefaultseppunct}\relax
\EndOfBibitem
\bibitem[Dai \latin{et~al.}(2018)Dai, Pavan~Kumar, Zhu, MacLachlan, Smith, and Wolf]{Dai}
Dai,~Y.; Pavan~Kumar,~V.; Zhu,~C.; MacLachlan,~M.~J.; Smith,~K.~J.; Wolf,~M.~O. Mesoporous Silica-Supported Nanostructured PdO/CeO2 Catalysts for Low-Temperature Methane Oxidation. \emph{ACS Applied Materials \& Interfaces} \textbf{2018}, \emph{10}, 477--487\relax
\mciteBstWouldAddEndPuncttrue
\mciteSetBstMidEndSepPunct{\mcitedefaultmidpunct}
{\mcitedefaultendpunct}{\mcitedefaultseppunct}\relax
\EndOfBibitem
\bibitem[Wang \latin{et~al.}(2024)Wang, Xu, Xiong, Li, Chen, Miao, Zhang, and Tang]{Wang}
Wang,~C.; Xu,~Y.; Xiong,~L.; Li,~X.; Chen,~E.; Miao,~T.~J.; Zhang,~Y.,~Tianyu~Lan; Tang,~J. Selective oxidation of methane to C2+ products over Au-CeO2 by photon-phonon co-driven catalysis. \emph{Nature Communications} \textbf{2024}, \emph{15}, 7535\relax
\mciteBstWouldAddEndPuncttrue
\mciteSetBstMidEndSepPunct{\mcitedefaultmidpunct}
{\mcitedefaultendpunct}{\mcitedefaultseppunct}\relax
\EndOfBibitem
\bibitem[Chen \latin{et~al.}(2024)Chen, Rana, Zhang, Hoffman, Huang, Yang, Vila, Perez-Aguilar, Hong, Li, Zeng, Chi, Kronawitter, Wang, Bare, Kulkarni, and Gates]{chen1}
Chen,~Y. \latin{et~al.}  Dynamic structural evolution of MgO-supported palladium catalysts: from metal to metal oxide nanoparticles to surface then subsurface atomically dispersed cations. \emph{Chem. Sci.} \textbf{2024}, \emph{15}, 6454--6464\relax
\mciteBstWouldAddEndPuncttrue
\mciteSetBstMidEndSepPunct{\mcitedefaultmidpunct}
{\mcitedefaultendpunct}{\mcitedefaultseppunct}\relax
\EndOfBibitem
\bibitem[Kim \latin{et~al.}(2018)Kim, Hwang, Jeong, Son, and Lim]{young}
Kim,~Y.~K.; Hwang,~S.-H.; Jeong,~S.~M.; Son,~K.~Y.; Lim,~S.~K. Colorimetric hydrogen gas sensor based on PdO/metal oxides hybrid nanoparticles. \emph{Talanta} \textbf{2018}, \emph{188}, 356--364\relax
\mciteBstWouldAddEndPuncttrue
\mciteSetBstMidEndSepPunct{\mcitedefaultmidpunct}
{\mcitedefaultendpunct}{\mcitedefaultseppunct}\relax
\EndOfBibitem
\bibitem[Mhlongo \latin{et~al.}(2019)Mhlongo, Motaung, Cummings, Swart, and Ray]{Mhlongo}
Mhlongo,~G.~H.; Motaung,~D.~E.; Cummings,~F.~R.; Swart,~H.~C.; Ray,~S.~S. A highly responsive NH3 sensor based on Pd-loaded ZnO nanoparticles prepared via a chemical precipitation approach. \emph{Scientific Reports} \textbf{2019}, \emph{9}, 9881\relax
\mciteBstWouldAddEndPuncttrue
\mciteSetBstMidEndSepPunct{\mcitedefaultmidpunct}
{\mcitedefaultendpunct}{\mcitedefaultseppunct}\relax
\EndOfBibitem
\bibitem[McCarty(1995)]{Jon}
McCarty,~J.~G. Kinetics of PdO combustion catalysis. \emph{Catalysis Today} \textbf{1995}, \emph{26}, 283--293, Selected papers presented at the International Workshop on Catalytic Combustion\relax
\mciteBstWouldAddEndPuncttrue
\mciteSetBstMidEndSepPunct{\mcitedefaultmidpunct}
{\mcitedefaultendpunct}{\mcitedefaultseppunct}\relax
\EndOfBibitem
\bibitem[Yu \latin{et~al.}(2024)Yu, Genz, Mendes, Ye, Meirer, Nachtegaal, Monai, and Weckhuysen]{xiang}
Yu,~X.; Genz,~N.~S.; Mendes,~R.~G.; Ye,~X.; Meirer,~F.; Nachtegaal,~M.; Monai,~M.; Weckhuysen,~B.~M. Anchoring PdOx clusters on defective alumina for improved catalytic methane oxidation. \emph{Nature Communications} \textbf{2024}, \emph{15}, 6494\relax
\mciteBstWouldAddEndPuncttrue
\mciteSetBstMidEndSepPunct{\mcitedefaultmidpunct}
{\mcitedefaultendpunct}{\mcitedefaultseppunct}\relax
\EndOfBibitem
\bibitem[Zheng \latin{et~al.}(2022)Zheng, Cao, Meng, Xiao, Ulstrup, Zhang, Zhao, Engelbrekt, and Xiao]{zheng}
Zheng,~Z.; Cao,~H.; Meng,~J.; Xiao,~Y.; Ulstrup,~J.; Zhang,~J.; Zhao,~F.; Engelbrekt,~C.; Xiao,~X. Synthesis and Structure of a Two-Dimensional Palladium Oxide Network on Reduced Graphene Oxide. \emph{Nano Letters} \textbf{2022}, \emph{22}, 4854--4860\relax
\mciteBstWouldAddEndPuncttrue
\mciteSetBstMidEndSepPunct{\mcitedefaultmidpunct}
{\mcitedefaultendpunct}{\mcitedefaultseppunct}\relax
\EndOfBibitem
\bibitem[Wang \latin{et~al.}(2024)Wang, Yu, Mu, Chen, Xiao, and Peng]{wang2}
Wang,~T.; Yu,~B.; Mu,~L.; Chen,~Z.; Xiao,~C.; Peng,~S. How to Characterize Supported PdOx Catalysts by CO-FTIR Spectroscopy: Importance of Surface Reduction and Particle Size. \emph{The Journal of Physical Chemistry C} \textbf{2024}, \emph{128}, 15356--15366\relax
\mciteBstWouldAddEndPuncttrue
\mciteSetBstMidEndSepPunct{\mcitedefaultmidpunct}
{\mcitedefaultendpunct}{\mcitedefaultseppunct}\relax
\EndOfBibitem
\bibitem[Jiang \latin{et~al.}(2020)Jiang, Khivantsev, and Wang]{Jiang2020}
Jiang,~D.; Khivantsev,~K.; Wang,~Y. Low-Temperature Methane Oxidation for Efficient Emission Control in Natural Gas Vehicles: Pd and Beyond. \emph{ACS Catalysis} \textbf{2020}, \emph{10}, 14304--14314\relax
\mciteBstWouldAddEndPuncttrue
\mciteSetBstMidEndSepPunct{\mcitedefaultmidpunct}
{\mcitedefaultendpunct}{\mcitedefaultseppunct}\relax
\EndOfBibitem
\bibitem[Bossche and Grönbeck(2015)Bossche, and Grönbeck]{bossche}
Bossche,~M. V.~d.; Grönbeck,~H. Methane Oxidation over PdO(101) Revealed by First-Principles Kinetic Modeling. \emph{Journal of the American Chemical Society} \textbf{2015}, \emph{137}, 12035--12044\relax
\mciteBstWouldAddEndPuncttrue
\mciteSetBstMidEndSepPunct{\mcitedefaultmidpunct}
{\mcitedefaultendpunct}{\mcitedefaultseppunct}\relax
\EndOfBibitem
\bibitem[Kulkarni \latin{et~al.}(2024)Kulkarni, Lezcano, Velisoju, Realpe, and Castaño]{Kulkarni}
Kulkarni,~S.~R.; Lezcano,~G.; Velisoju,~V.~K.; Realpe,~N.; Castaño,~P. Microkinetic Modeling to Decode Catalytic Reactions and Empower Catalytic Design. \emph{ChemCatChem} \textbf{2024}, \emph{16}, e202301720\relax
\mciteBstWouldAddEndPuncttrue
\mciteSetBstMidEndSepPunct{\mcitedefaultmidpunct}
{\mcitedefaultendpunct}{\mcitedefaultseppunct}\relax
\EndOfBibitem
\bibitem[Motagamwala and Dumesic(2021)Motagamwala, and Dumesic]{Motagamwala}
Motagamwala,~A.~H.; Dumesic,~J.~A. Microkinetic Modeling: A Tool for Rational Catalyst Design. \emph{Chemical Reviews} \textbf{2021}, \emph{121}, 1049--1076\relax
\mciteBstWouldAddEndPuncttrue
\mciteSetBstMidEndSepPunct{\mcitedefaultmidpunct}
{\mcitedefaultendpunct}{\mcitedefaultseppunct}\relax
\EndOfBibitem
\bibitem[Keller \latin{et~al.}(2020)Keller, Lott, Stotz, Maier, and Deutschmann]{keller}
Keller,~K.; Lott,~P.; Stotz,~H.; Maier,~L.; Deutschmann,~O. Microkinetic Modeling of the Oxidation of Methane Over PdO Catalysts—Towards a Better Understanding of the Water Inhibition Effect. \emph{Catalysts} \textbf{2020}, \emph{10}\relax
\mciteBstWouldAddEndPuncttrue
\mciteSetBstMidEndSepPunct{\mcitedefaultmidpunct}
{\mcitedefaultendpunct}{\mcitedefaultseppunct}\relax
\EndOfBibitem
\bibitem[Gupta \latin{et~al.}(2024)Gupta, Tomar, Choi, Jeong, Lee, and Bhattacharjee]{SGupta}
Gupta,~S.; Tomar,~S.; Choi,~J.~H.; Jeong,~H.; Lee,~S.-C.; Bhattacharjee,~S. Controlling Moisture for Enhanced Ozone Decomposition: A Study of Water Effects on CeO2 Surfaces and Catalytic Activity. \emph{The Journal of Physical Chemistry C} \textbf{2024}, \emph{128}, 5889--5899\relax
\mciteBstWouldAddEndPuncttrue
\mciteSetBstMidEndSepPunct{\mcitedefaultmidpunct}
{\mcitedefaultendpunct}{\mcitedefaultseppunct}\relax
\EndOfBibitem
\bibitem[Colussi \latin{et~al.}(2020)Colussi, Fornasiero, and Trovarelli]{sara202}
Colussi,~S.; Fornasiero,~P.; Trovarelli,~A. Structure-activity relationship in Pd/CeO2 methane oxidation catalysts. \emph{Chinese Journal of Catalysis} \textbf{2020}, \emph{41}, 938--950\relax
\mciteBstWouldAddEndPuncttrue
\mciteSetBstMidEndSepPunct{\mcitedefaultmidpunct}
{\mcitedefaultendpunct}{\mcitedefaultseppunct}\relax
\EndOfBibitem
\bibitem[Hohenberg and Kohn(1964)Hohenberg, and Kohn]{DFT1}
Hohenberg,~P.; Kohn,~W. Inhomogeneous Electron Gas. \emph{Phys. Rev.} \textbf{1964}, \emph{136}, B864--B871\relax
\mciteBstWouldAddEndPuncttrue
\mciteSetBstMidEndSepPunct{\mcitedefaultmidpunct}
{\mcitedefaultendpunct}{\mcitedefaultseppunct}\relax
\EndOfBibitem
\bibitem[Kohn and Sham(1965)Kohn, and Sham]{DFT2}
Kohn,~W.; Sham,~L.~J. Self-Consistent Equations Including Exchange and Correlation Effects. \emph{Phys. Rev.} \textbf{1965}, \emph{140}, A1133--A1138\relax
\mciteBstWouldAddEndPuncttrue
\mciteSetBstMidEndSepPunct{\mcitedefaultmidpunct}
{\mcitedefaultendpunct}{\mcitedefaultseppunct}\relax
\EndOfBibitem
\bibitem[Aouina \latin{et~al.}(2023)Aouina, Gatti, Chen, Zhang, and Reining]{kohnsham}
Aouina,~A.; Gatti,~M.; Chen,~S.; Zhang,~S.; Reining,~L. Accurate Kohn-Sham auxiliary system from the ground-state density of solids. \emph{Phys. Rev. B} \textbf{2023}, \emph{107}, 195123\relax
\mciteBstWouldAddEndPuncttrue
\mciteSetBstMidEndSepPunct{\mcitedefaultmidpunct}
{\mcitedefaultendpunct}{\mcitedefaultseppunct}\relax
\EndOfBibitem
\bibitem[Kresse and Furthm\"uller(1996)Kresse, and Furthm\"uller]{VASP1}
Kresse,~G.; Furthm\"uller,~J. Efficient iterative schemes for ab initio total-energy calculations using a plane-wave basis set. \emph{Phys. Rev. B} \textbf{1996}, \emph{54}, 11169--11186\relax
\mciteBstWouldAddEndPuncttrue
\mciteSetBstMidEndSepPunct{\mcitedefaultmidpunct}
{\mcitedefaultendpunct}{\mcitedefaultseppunct}\relax
\EndOfBibitem
\bibitem[Kresse and Joubert(1999)Kresse, and Joubert]{PAW}
Kresse,~G.; Joubert,~D. From ultrasoft pseudopotentials to the projector augmented-wave method. \emph{Phys. Rev. B} \textbf{1999}, \emph{59}, 1758--1775\relax
\mciteBstWouldAddEndPuncttrue
\mciteSetBstMidEndSepPunct{\mcitedefaultmidpunct}
{\mcitedefaultendpunct}{\mcitedefaultseppunct}\relax
\EndOfBibitem
\bibitem[Dudarev \latin{et~al.}(1998)Dudarev, Botton, Savrasov, Humphreys, and Sutton]{Dudarev}
Dudarev,~S.~L.; Botton,~G.~A.; Savrasov,~S.~Y.; Humphreys,~C.~J.; Sutton,~A.~P. Electron-energy-loss spectra and the structural stability of nickel oxide: An LSDA+U study. \emph{Phys. Rev. B} \textbf{1998}, \emph{57}, 1505--1509\relax
\mciteBstWouldAddEndPuncttrue
\mciteSetBstMidEndSepPunct{\mcitedefaultmidpunct}
{\mcitedefaultendpunct}{\mcitedefaultseppunct}\relax
\EndOfBibitem
\bibitem[Grimme \latin{et~al.}(2010)Grimme, Antony, Ehrlich, and Krieg]{DFTD}
Grimme,~S.; Antony,~J.; Ehrlich,~S.; Krieg,~H. A consistent and accurate ab initio parametrization of density functional dispersion correction (DFT-D) for the 94 elements H-Pu. \emph{The Journal of Chemical Physics} \textbf{2010}, \emph{132}, 154104\relax
\mciteBstWouldAddEndPuncttrue
\mciteSetBstMidEndSepPunct{\mcitedefaultmidpunct}
{\mcitedefaultendpunct}{\mcitedefaultseppunct}\relax
\EndOfBibitem
\bibitem[Sanville \latin{et~al.}(2007)Sanville, Kenny, Smith, and Henkelman]{bader}
Sanville,~E.; Kenny,~S.~D.; Smith,~R.; Henkelman,~G. Improved grid-based algorithm for Bader charge allocation. \emph{Journal of Computational Chemistry} \textbf{2007}, \emph{28}, 899--908\relax
\mciteBstWouldAddEndPuncttrue
\mciteSetBstMidEndSepPunct{\mcitedefaultmidpunct}
{\mcitedefaultendpunct}{\mcitedefaultseppunct}\relax
\EndOfBibitem
\bibitem[Penschke \latin{et~al.}(2013)Penschke, Paier, and Sauer]{penschke}
Penschke,~C.; Paier,~J.; Sauer,~J. Oligomeric Vanadium Oxide Species Supported on the CeO2(111) Surface: Structure and Reactivity Studied by Density Functional Theory. \emph{The Journal of Physical Chemistry C} \textbf{2013}, \emph{117}, 5274--5285\relax
\mciteBstWouldAddEndPuncttrue
\mciteSetBstMidEndSepPunct{\mcitedefaultmidpunct}
{\mcitedefaultendpunct}{\mcitedefaultseppunct}\relax
\EndOfBibitem
\bibitem[Lawler \latin{et~al.}(2020)Lawler, Cho, Ham, Ju, Lee, Kim, Il~Choi, and Jang]{lawler}
Lawler,~R.; Cho,~J.; Ham,~H.~C.; Ju,~H.; Lee,~S.~W.; Kim,~J.~Y.; Il~Choi,~J.; Jang,~S.~S. CeO2(111) Surface with Oxygen Vacancy for Radical Scavenging: A Density Functional Theory Approach. \emph{The Journal of Physical Chemistry C} \textbf{2020}, \emph{124}, 20950--20959\relax
\mciteBstWouldAddEndPuncttrue
\mciteSetBstMidEndSepPunct{\mcitedefaultmidpunct}
{\mcitedefaultendpunct}{\mcitedefaultseppunct}\relax
\EndOfBibitem
\bibitem[Fan \latin{et~al.}(2016)Fan, Li, Zhao, Shan, and Xu]{Fan}
Fan,~J.; Li,~C.; Zhao,~J.; Shan,~Y.; Xu,~H. The Enhancement of Surface Reactivity on CeO2 (111) Mediated by Subsurface Oxygen Vacancies. \emph{The Journal of Physical Chemistry C} \textbf{2016}, \emph{120}, 27917--27924\relax
\mciteBstWouldAddEndPuncttrue
\mciteSetBstMidEndSepPunct{\mcitedefaultmidpunct}
{\mcitedefaultendpunct}{\mcitedefaultseppunct}\relax
\EndOfBibitem
\bibitem[Tomar \latin{et~al.}(2024)Tomar, Bhadoria, Jeong, Choi, Lee, and Bhattacharjee]{tomar}
Tomar,~S.; Bhadoria,~B.~S.; Jeong,~H.; Choi,~J.~H.; Lee,~S.-C.; Bhattacharjee,~S. Single-Atom Pd Catalyst on a CeO2 (111) Surface for Methane Oxidation: Activation Barriers and Reaction Pathways. \emph{The Journal of Physical Chemistry C} \textbf{2024}, \emph{128}, 8580--8589\relax
\mciteBstWouldAddEndPuncttrue
\mciteSetBstMidEndSepPunct{\mcitedefaultmidpunct}
{\mcitedefaultendpunct}{\mcitedefaultseppunct}\relax
\EndOfBibitem
\bibitem[Henkelman and Jónsson(2000)Henkelman, and Jónsson]{NEB}
Henkelman,~G.; Jónsson,~H. {Improved tangent estimate in the nudged elastic band method for finding minimum energy paths and saddle points}. \emph{The Journal of Chemical Physics} \textbf{2000}, \emph{113}, 9978--9985\relax
\mciteBstWouldAddEndPuncttrue
\mciteSetBstMidEndSepPunct{\mcitedefaultmidpunct}
{\mcitedefaultendpunct}{\mcitedefaultseppunct}\relax
\EndOfBibitem
\bibitem[Filot \latin{et~al.}(2014)Filot, vanSanten, and Hensen]{Filot}
Filot,~I. A.~W.; vanSanten,~R.~A.; Hensen,~E. J.~M. The Optimally Performing Fischer–Tropsch Catalyst. \emph{Angewandte Chemie International Edition} \textbf{2014}, \emph{53}, 12746--12750\relax
\mciteBstWouldAddEndPuncttrue
\mciteSetBstMidEndSepPunct{\mcitedefaultmidpunct}
{\mcitedefaultendpunct}{\mcitedefaultseppunct}\relax
\EndOfBibitem
\bibitem[McQuarrie and D.(1999)McQuarrie, and D.]{donald}
McQuarrie,~D.~A.; D.,~S.~J. \emph{{M}olecular {T}hermodynamics}; University Science Books: California, 1999\relax
\mciteBstWouldAddEndPuncttrue
\mciteSetBstMidEndSepPunct{\mcitedefaultmidpunct}
{\mcitedefaultendpunct}{\mcitedefaultseppunct}\relax
\EndOfBibitem
\bibitem[Atkins and Paula(2018)Atkins, and Paula]{atkins}
Atkins,~J. K.~P.; Paula,~J.~d. \emph{{A}tkins {P}hysical {C}hemistry}; Oxford University Press: Oxford, 2018\relax
\mciteBstWouldAddEndPuncttrue
\mciteSetBstMidEndSepPunct{\mcitedefaultmidpunct}
{\mcitedefaultendpunct}{\mcitedefaultseppunct}\relax
\EndOfBibitem
\bibitem[Trinchero \latin{et~al.}(2013)Trinchero, Hellman, and Grönbeck]{adriana}
Trinchero,~A.; Hellman,~A.; Grönbeck,~H. Methane oxidation over Pd and Pt studied by DFT and kinetic modeling. \emph{Surface Science} \textbf{2013}, \emph{616}, 206--213\relax
\mciteBstWouldAddEndPuncttrue
\mciteSetBstMidEndSepPunct{\mcitedefaultmidpunct}
{\mcitedefaultendpunct}{\mcitedefaultseppunct}\relax
\EndOfBibitem
\bibitem[Jiao and Wang(2024)Jiao, and Wang]{Jiao}
Jiao,~H.; Wang,~G.-C. Dry Reforming of Methane on Ni/LaZrO2 Catalyst under External Electric Fields: A Combined First-Principles and Microkinetic Modeling Study. \emph{ACS Applied Materials \& Interfaces} \textbf{2024}, \emph{16}, 35166--35178\relax
\mciteBstWouldAddEndPuncttrue
\mciteSetBstMidEndSepPunct{\mcitedefaultmidpunct}
{\mcitedefaultendpunct}{\mcitedefaultseppunct}\relax
\EndOfBibitem
\bibitem[Deng \latin{et~al.}(2020)Deng, Song, Jing, Yu, Zhao, Xu, and Liu]{Jianlin}
Deng,~J.; Song,~W.; Jing,~M.; Yu,~T.; Zhao,~Z.; Xu,~C.; Liu,~J. A DFT and microkinetic study of HCHO catalytic oxidation mechanism over Pd/Co3O4 catalysts: The effect of metal-oxide interface. \emph{Catalysis Today} \textbf{2020}, \emph{339}, 210--219\relax
\mciteBstWouldAddEndPuncttrue
\mciteSetBstMidEndSepPunct{\mcitedefaultmidpunct}
{\mcitedefaultendpunct}{\mcitedefaultseppunct}\relax
\EndOfBibitem
\bibitem[Moraes \latin{et~al.}(2023)Moraes, Bittencourt, Andriani, and Da~Silva]{Moraes}
Moraes,~P. I.~R.; Bittencourt,~A. F.~B.; Andriani,~K.~F.; Da~Silva,~J. L.~F. Theoretical Insights into Methane Activation on Transition-Metal Single-Atom Catalysts Supported on the CeO2(111) Surface. \emph{The Journal of Physical Chemistry C} \textbf{2023}, \emph{127}, 16357--16366\relax
\mciteBstWouldAddEndPuncttrue
\mciteSetBstMidEndSepPunct{\mcitedefaultmidpunct}
{\mcitedefaultendpunct}{\mcitedefaultseppunct}\relax
\EndOfBibitem
\bibitem[Spezzati \latin{et~al.}(2017)Spezzati, Su, Hofmann, Benavidez, DeLaRiva, McCabe, Datye, and Hensen]{spezzati}
Spezzati,~G.; Su,~Y.; Hofmann,~J.~P.; Benavidez,~A.~D.; DeLaRiva,~A.~T.; McCabe,~J.; Datye,~A.~K.; Hensen,~E. J.~M. Atomically Dispersed Pd–O Species on CeO2(111) as Highly Active Sites for Low-Temperature CO Oxidation. \emph{ACS Catalysis} \textbf{2017}, \emph{7}, 6887--6891\relax
\mciteBstWouldAddEndPuncttrue
\mciteSetBstMidEndSepPunct{\mcitedefaultmidpunct}
{\mcitedefaultendpunct}{\mcitedefaultseppunct}\relax
\EndOfBibitem
\bibitem[Zhang \latin{et~al.}(2023)Zhang, Yang, Qin, Hu, Zhao, Zhao, Cao, Gao, Zhou, Liang, Tan, and Qin]{zhang}
Zhang,~J.; Yang,~Y.; Qin,~F.; Hu,~T.; Zhao,~X.; Zhao,~S.; Cao,~Y.; Gao,~Z.; Zhou,~Z.; Liang,~R.; Tan,~C.; Qin,~Y. Catalyzing Generation and Stabilization of Oxygen Vacancies on CeO2x Nanorods by Pt Nanoclusters as Nanozymes for Catalytic Therapy. \emph{Advanced Healthcare Materials} \textbf{2023}, \emph{12}, 2302056\relax
\mciteBstWouldAddEndPuncttrue
\mciteSetBstMidEndSepPunct{\mcitedefaultmidpunct}
{\mcitedefaultendpunct}{\mcitedefaultseppunct}\relax
\EndOfBibitem
\bibitem[Yu \latin{et~al.}(2022)Yu, Cheng, Wang, Xiao, Xing, Ren, Lu, Li, Zhuang, and Chen]{shiming}
Yu,~S.; Cheng,~X.; Wang,~Y.; Xiao,~B.; Xing,~Y.; Ren,~J.; Lu,~Y.; Li,~H.; Zhuang,~C.; Chen,~G. High activity and selectivity of single palladium atom for oxygen hydrogenation to H2O2. \emph{Nature Communications} \textbf{2022}, \emph{13}, 4737\relax
\mciteBstWouldAddEndPuncttrue
\mciteSetBstMidEndSepPunct{\mcitedefaultmidpunct}
{\mcitedefaultendpunct}{\mcitedefaultseppunct}\relax
\EndOfBibitem
\bibitem[Ouyang \latin{et~al.}(2018)Ouyang, Curtarolo, Ahmetcik, Scheffler, and Ghiringhelli]{ouyang2018sisso}
Ouyang,~R.; Curtarolo,~S.; Ahmetcik,~E.; Scheffler,~M.; Ghiringhelli,~L.~M. SISSO: A compressed-sensing method for identifying the best low-dimensional descriptor in an immensity of offered candidates. \emph{Physical Review Materials} \textbf{2018}, \emph{2}, 083802\relax
\mciteBstWouldAddEndPuncttrue
\mciteSetBstMidEndSepPunct{\mcitedefaultmidpunct}
{\mcitedefaultendpunct}{\mcitedefaultseppunct}\relax
\EndOfBibitem
\bibitem[Bhattacharjee \latin{et~al.}(2016)Bhattacharjee, Yoo, Waghmare, and Lee]{bhattacharjee2016nh3}
Bhattacharjee,~S.; Yoo,~S.; Waghmare,~U.~V.; Lee,~S. NH3 adsorption on PtM (Fe, Co, Ni) surfaces: Cooperating effects of charge transfer, magnetic ordering and lattice strain. \emph{Chemical Physics Letters} \textbf{2016}, \emph{648}, 166--169\relax
\mciteBstWouldAddEndPuncttrue
\mciteSetBstMidEndSepPunct{\mcitedefaultmidpunct}
{\mcitedefaultendpunct}{\mcitedefaultseppunct}\relax
\EndOfBibitem
\bibitem[Ouyang \latin{et~al.}(2019)Ouyang, Ahmetcik, Carbogno, Scheffler, and Ghiringhelli]{ouyang2019simultaneous}
Ouyang,~R.; Ahmetcik,~E.; Carbogno,~C.; Scheffler,~M.; Ghiringhelli,~L.~M. Simultaneous learning of several materials properties from incomplete databases with multi-task SISSO. \emph{Journal of Physics: Materials} \textbf{2019}, \emph{2}, 024002\relax
\mciteBstWouldAddEndPuncttrue
\mciteSetBstMidEndSepPunct{\mcitedefaultmidpunct}
{\mcitedefaultendpunct}{\mcitedefaultseppunct}\relax
\EndOfBibitem
\end{mcitethebibliography}

\end{document}